\documentclass[fleqn]{2023SCGE}
\usepackage[dvipsnames]{xcolor} 
\usepackage{orcidlink} 
\usepackage{ulem}

\begin{document}

\ensubject{subject}

\ArticleType{Article}
\SpecialTopic{SPECIAL TOPIC: }
\Year{2025}
\Month{January}
\Vol{66}
\No{1}
\DOI{??}
\ArtNo{000000}
\ReceiveDate{February 16, 2025}
\AcceptDate{*****}
\renewcommand\floatpagefraction{.9}
\renewcommand\topfraction{.9}
\renewcommand\bottomfraction{.9}
\renewcommand\textfraction{.1}
\setcounter{totalnumber}{50}
\setcounter{topnumber}{50}
\setcounter{bottomnumber}{50}

\newcommand{\red}{\textcolor{red} }
\newcommand{\green}{\textcolor{green} }
\newcommand{\orange}{\textcolor{orange}}
\newcommand{\cyan}{\textcolor{cyan}}
\newcommand{\magenta}{\textcolor{magenta}}

\newcommand{\be}{\begin{equation}}
\newcommand{\ee}{\end{equation}}
\newcommand{\ba}{\begin{eqnarray}}
\newcommand{\ea}{\end{eqnarray}}
\newcommand{\no}{\nonumber}
\newcommand{\bi}{\begin{itemize}}
\newcommand{\ei}{\end{itemize}}
\newcommand{\kpch}{h^{-1} {\rm kpc}}
\newcommand{\mpch}{h^{-1} {\rm Mpc}}
\newcommand{\mpcht}{h^{-3} {\rm Mpc^3}}
\newcommand{\gpch}{h^{-1} {\rm Gpc}}
\newcommand{\gpcht}{h^{-3} {\rm Gpc^3}}
\newcommand{\hmpc}{h {\rm Mpc}^{-1}}
\newcommand{\hmpct}{h^3 {\rm Mpc}^{-3}}
\newcommand{\hgpc}{h{\rm Gpc}^{-1} }
\newcommand{\hgpct}{h^3 {\rm Gpc}^{-3} }
\newcommand{\rmnum}[1]{\romannumeral #1}
\newcommand{\Rmnum}[1]{\uppercase\expandafter{\romannumeral #1}}
\newcommand{\Msun}{M_{\odot}}

\title{CSST Cosmological Emulator I: Matter Power Spectrum Emulation with one percent accuracy to $k=10\,h/\mathrm{Mpc}$}{CSST Cosmological Emulator I}

\author[1,2,3]{Zhao Chen\orcidlink{0000-0002-2183-9863}}{} 
\author[1,2,3]{Yu Yu\orcidlink{0000-0002-9359-7170}}{{}}
\author[1,2,3]{Jiaxin Han\orcidlink{0000-0002-8010-6715}}{}
\author[1,2,4,3]{Yipeng Jing\orcidlink{0000-0002-4534-3125}}{}%

\thanks{Corresponding author (email:~\href{yuyu22@sjtu.edu.cn}{yuyu22@sjtu.edu.cn})}

\AuthorMark{Chen Zhao}

\AuthorCitation{Chen Z, Yu Y, et al.}

\address[1]{Department of Astronomy, School of Physics and Astronomy, Shanghai Jiao Tong University, Shanghai 200240, China}
\address[2]{State Key Laboratory of Dark Matter Physics, School of Physics and Astronomy, Shanghai Jiao Tong University, Shanghai 200240, China}
\address[3]{Key Laboratory for Particle Astrophysics and Cosmology (MOE)/Shanghai Key Laboratory for Particle Physics and Cosmology, Shanghai 200240, China}
\address[4]{Tsung-Dao Lee Institute, Shanghai Jiao Tong University, Shanghai 201210, China}


\abstract{
In the near future, the China Space Station Telescope (CSST) will obtain unprecedented imaging and spectroscopic data.
The statistical errors in the cosmological parameter constraints will be reduced significantly.
The corresponding theoretical tools must meet the percent-level accuracy required to extract as much cosmological information as possible from the observations.
We present the \texttt{CSST Emulator} to provide nonlinear power spectrum predictions in the eight cosmological parameter space $\Omega_\mathrm{cb},\Omega_\mathrm{b},H_{0},n_{s},A_{s},w_{0}, w_{a}$, and $m_\nu$.
It is constructed based on the \textsc{Kun} simulation suite, consisting of 129 high-resolution simulations with box size $L=1\,\gpch$ and evolving $3072^3$ particles.
The determinations of parameter ranges, sampling method, and emulation strategy in the whole construction have been optimized exquisitely.
This enables our prediction for $k\leq 10\,\hmpc$ and $z\leq 2.0$ to reach $1\%$ accuracy validated through internal and external simulations.
We also compare our results with recent \texttt{BACCO}, \texttt{EuclidEmulator2}, and  \texttt{Mira-Titan IV} emulators, which demonstrate the \texttt{CSST Emulator}'s excellent performance across a wide cosmological parameter range in the nonlinear regime.
\texttt{CSST Emulator} is publicly released at \url{https://github.com/czymh/csstemu}, and provides a fundamental theoretical tool for accurate cosmological inference with future CSST observations.
}



\maketitle


\begin{multicols}{2}

\section{Introduction}
\label{sec:intro}

Since the exquisite observations of the cosmic microwave background (CMB) over several decades, the standard cosmological model with Einstein's cosmological constant ($\Lambda\mathrm{CDM}$) has become the baseline model to extend our knowledge of the nature and evolution of our Universe.
Currently, it is commonly acknowledged that dark matter and dark energy constitute the principal components of the Universe.
However, this early-time probe only provides limited constraints on the nature of the two dark components.
The study of large-scale structure (LSS) at late times over the past billion years can provide complementary constraints.

The Stage-\Rmnum{3} optical cosmological experiments, including the Dark Energy Survey (DES\footnote{\url{http://www.darkenergysurvey.org/}} \cite{2005astro.ph.10346T}), the Hyper Suprime-Cam Subaru Strategic Program (HSC\footnote{\url{https://www.naoj.org/Projects/HSC/}} \cite{2018PASJ...70S...4A}), the Kilo-Degree Survey (KiDS\footnote{\url{http://kids.strw.leidenuniv.nl/}} \cite{2013Msngr.154...44D,2013ExA....35...25D}) and the Sloan Digital Sky Survey (SDSS\footnote{\url{https://www.sdss.org/}} \cite{2000AJ....120.1579Y}),
have measured the cosmic geometry and the structure growth accurately.
With the development of nonlinear theoretical predictions, these imaging and spectroscopic surveys can constrain the parameters under the $\Lambda\mathrm{CDM}$ cosmology, and even beyond.
The ongoing and forthcoming Stage-\Rmnum{4} galaxy surveys, such as the Dark Energy Spectroscopic Instrument (DESI\footnote{\url{https://www.desi.lbl.gov/}} \cite{2016arXiv161100036D}), the Vera Rubin Observatory Legacy Survey of Space and Time (LSST\footnote{\url{http://www.lsst.org}} \cite{2009arXiv0912.0201L}), the Euclid satellite\footnote{\url{http://www.euclid-ec.org}}~\cite{2011arXiv1110.3193L,2024arXiv240513491E}, the Nancy Grace Roman Space Telescope (Roman\footnote{\url{https://roman.gsfc.nasa.gov/}} \cite{2019BAAS...51c.341D}), and the China Space Station Telescope (CSST\footnote{\url{https://www.nao.cas.cn/csst/}} \cite{2019ApJ...883..203G}) will improve our understanding about the nature of dark energy, enable accurate measurement of the total neutrino mass, and even have the potential to solve the neutrino mass hierarchy problem (e.g., \cite{2024arXiv240513491E}).

However, several puzzles in the standard cosmological model have appeared with the improvement in the accuracy of observations (see \cite{2022NewAR..9501659P} for a review).
For example, the local distance-ladder Hubble constant $H_\mathrm{0}$ measurements are significantly higher ($>4\sigma$) than the results from the CMB fluctuations in the context of $\Lambda\mathrm{CDM}$ cosmology, which is referred to as the Hubble tension (e.g., \cite{2021CQGra..38o3001D,2021MNRAS.502.2065D,2021A&ARv..29....9S,2022ApJ...934L...7R}).
The growth tension refers to the $2\sim3\sigma$ discrepancy between the growth rate measured from the late-time probes (such as weak gravitational lensing and redshift-space distortions) and values inferred from the Planck cosmology \cite{2018PhRvD..98d3526A,2022PhRvD.105b3514A,2022ARNPS..72....1T}.
The recent DESI baryon acoustic oscillations (BAO) measurements in the galaxy, quasar, and Lyman-$\alpha$ forest tracers prefer the universe with $w_{0} >-1$ and $w_{a}<0$ under the Chevallier-Polarski-Linder (CPL \cite{2001IJMPD..10..213C,2003PhRvL..90i1301L}) dynamic dark energy model when combined with CMB or supernova Type Ia (SNIa) data \cite{2024arXiv240403002D}.
These challenges may point to new physics or potential observational and theoretical systematic errors.

Due to the complicated late-time nonlinear evolution, the most accurate prediction on small scales relies on expensive cosmological simulations.
However, it is numerically impossible to directly utilize this method for Bayesian parameter inference.
The first solution for the nonlinear theory is the parametric fitting approach based on the halo model (see \cite{2023OJAp....6E..39A} for the recent review).
The original HaloFit model was proposed for fitting nonlinear power spectra of various cosmological models in \cite{2003MNRAS.341.1311S}.
This successful tool was revisited by using 16 modern cold dark matter simulations in \cite{2012ApJ...761..152T} and extended for cosmology with massive neutrinos in \cite{2012MNRAS.420.2551B}.
Smith \& Angulo 2019 \cite{2019MNRAS.486.1448S} updated this to NGENHaloFit by combining perturbation theory on large scales and correcting the difference between revised HaloFit and their simulation results for improved accuracy.
Mead et al.~2015 \cite{2015MNRAS.454.1958M} presented a new halo-model formalism with fewer but physically motivated parameters to predict the nonlinear power spectrum with baryonic feedback.
Massive neutrinos, dynamic dark energy, and modified gravity models are included in  \cite{2016MNRAS.459.1468M}.
Later, it was further developed to provide more accurate predictions with the help of \texttt{FrankenEmu} \cite{2014ApJ...780..111H}.
However, this is still inadequate for the weak lensing study of Stage-\Rmnum{4}  surveys \cite{2023MNRAS.522.3766T}.

Another roadway is the `emulator' approach, which directly interpolates statistics measured from simulations in a given parameter space, without assuming a specific functional form for parameter dependence.
The first released simulation-based emulator, \texttt{CosmicEmu}, was proposed based on the \texttt{Coyote} Universe project in \cite{2010ApJ...715..104H,2009ApJ...705..156H,2010ApJ...713.1322L}, which provides percent-level accurate matter power spectrum at $k\lesssim 1\,\hmpc$ and $z\leq1$ for five varied cosmological parameters ($\omega_\mathrm{m},\omega_\mathrm{b},n_{s},\sigma_{8}$, and $w$).
This was then updated to \texttt{FrankenEmu} \cite{2014ApJ...780..111H} with extended scales and redshifts up to $k = 8.6\,\hmpc$ and $z=4$, and an extra independent variable, dimensionless Hubble parameter $h$.
However, these modifications produced worse accuracy ($\lesssim 5\%$) on the overall parameter space.
Casarini et al.~2016 \cite{2016JCAP...08..008C} introduced an efficient approach to extend the constant dark energy ($w$) to the CPL ($w_{0}, w_{a}$) parametrization.
To incorporate the effect of massive neutrinos for the near future surveys, the \texttt{Coyote} team further updated the \texttt{CosmicEmu} by training the new \texttt{Mira-Titan} Universe simulations \cite{2016ApJ...820..108H,2017ApJ...847...50L,2020ApJ...901....5B,2023MNRAS.520.3443M}, which include 111 cosmologies for eight cosmological parameters $( \omega_\mathrm{m}, \omega_\mathrm{b}, \sigma_{8}, h, n_{s}, w_{0}, w_{a}, \omega_\nu )$.
This new version quoted $2\sim3\%$ accuracy over their training sample space for $k\leq5\,\hmpc$ and $z\leq 2.02$ (hereafter  \texttt{Mira-Titan IV}).
Additionally, Angulo et al.~2021 \cite{2021MNRAS.507.5869A} constructed the \texttt{BACCO} emulator within an eight-dimensional parameter space ($\Omega_\mathrm{m},\Omega_\mathrm{b},n_{s},\sigma_{8},h,m_\nu, w_{0}, w_{a}$) for the power spectrum in the redshift range $z\in [0,\,1.5]$ and wavenumber range $k\in[0.01,\,5]\,\hmpc$. This was achieved using only three dark-matter-only simulations, leveraging the cosmology-rescaling technique \cite{2010MNRAS.405..143A,2019MNRAS.489.5938Z}. 
Emulators are commonly trained by the combination of principal component analysis (PCA) and Gaussian process regression (GPR).
In 2019, the \texttt{EuclidEmulator} \cite{2019MNRAS.484.5509E} was first released for the matter power spectrum prediction under the $w\mathrm{CDM}$ cosmology through the polynomial chaos expansion (PCE) instead of GPR.
It was then developed into the \texttt{EuclidEmulator2} \cite{2021MNRAS.505.2840E} with the inclusion of $m_\nu$ and $w_{a}$.
Recently, nonlinear power prediction using different emulation strategies (e.g., \cite{Agarwal12,Agarwal14,2022MNRAS.509.2551H,2023MNRAS.526.2903H}) and symbolic regression algorithms (e.g., \cite{2024A&A...686A.150B,2024arXiv241014623S}) has also been investigated, including extensions to modified gravity (e.g., \cite{2019PhRvD.100l3540W,2021PhRvD.103l3525R,2022MNRAS.515.4161A,2022JCAP...09..051B,2023JCAP...12..045F,2024A&A...685A.156M,2024MNRAS.527.7242S,2024ApJ...971...11B}).
Except for the fundamental two-point statistic of dark matter, there are other emulators for different compressed statistics, such as halo mass function (e.g., \cite{2019ApJ...872...53M,2020ApJ...901....5B,2024A&A...691A.323S,2024arXiv241000913S}), halo clustering (e.g., \cite{2019ApJ...884...29N,2020PhRvD.102f3504K}), galaxy clustering (e.g., \cite{2015ApJ...810...35K,2019ApJ...874...95Z,2020MNRAS.492.2872W,2022MNRAS.515..871Y,2023ApJ...948...99Z,2024ApJ...961..208S}), the basis of the biased expansion model (e.g., \cite{2021JCAP...09..020H,2023MNRAS.524.2407Z,2023MNRAS.520.3725P,2023JCAP...07..054D}), and other high-order statistics (e.g., \cite{2018JCAP...03..049L,2019PhRvD..99h3508L,2019PhRvD..99f3527L,2019JCAP...05..043C}).

In this work, we describe \textsc{Kun}, a new suite of $N$-body simulations including 129 different cosmologies under the $w_0 w_a \mathrm{CDM} + \sum m_{\nu}$ model, which is a part of \textsc{Jiutian} simulation suite \cite{Han2025} for the preparation of the upcoming CSST survey.
A key science goal of CSST is the weak lensing survey, which induces us to generate this new simulation suite with continuous particle light-cones to support the lensing peak analysis and field-level inference.
As the first application of this project, we construct the \texttt{CSST Emulator} with exquisite sampling and emulation design to provide $\sim1\%$ level of accuracy for power spectrum of the total matter and cold component (cold dark matter and baryon, hereafter referred as cb) up to $k_\mathrm{max} = 10\,\hmpc$ and $z\in [0,\,3]$.
At this limiting scale, baryonic effects, such as cooling and feedback, indeed affect the matter clustering.
Considering this complexity as extra dimensions will rapidly increase computational costs.
Fortunately, the influence on matter clustering can be addressed through other efficient approaches after the accomplishment of the emulator (e.g., \cite{2015JCAP...12..049S,2019JCAP...03..020S,2021MNRAS.506.4070A,2021JCAP...12..046G,2023MNRAS.523.2247S}).
Thus, we focus on the emulation of gravity-only simulations here.

The rest of this paper is structured as follows:
In Section~\ref{sec:design}, we describe the detailed design of the training samples in the $w_0 w_a \mathrm{CDM} + \sum m_{\nu}$ framework.
A surrogate emulator is constructed to determine the range of cosmological parameters and to optimize the sampling technique.
An overview of the \textsc{Kun} simulation suite is provided in Section~\ref{sec:simu}.
We also illustrate the special neutrino treatment and convergence test on simulation nuisance parameters in this part.
In Section~\ref{sec:emulation}, we present the smoothing preprocess and investigate the performance of different emulation strategies.
To validate our emulator, we perform various examinations by employing external simulations and comparing our results with previous emulators carefully in Section~\ref{sec:validation}.
In Section~\ref{sec:conc}, we summarize our overall emulator construction and provide possible future extensions of this released version.

\section{Parameter Space Design}
\label{sec:design}

The first step in constructing the emulator is to design a reasonable cosmological parameter space, which is crucial for broad application and precision prediction.
Taking into account the deviation between the cosmological constant model and the recent DESI 2024 observation \cite{2024arXiv240403002D}, it is essential to incorporate the dynamic dark energy model.
The constraint on the massive neutrino mass is a key goal of the Stage-\Rmnum{4} surveys (e.g., \cite{2024arXiv240506047E}).
Therefore, we consider the eight-dimensional parameter space, also referred to as the $w_0 w_a \mathrm{CDM} + \sum m_{\nu}$ model:
\begin{itemize}
    \item [i]   $\Omega_\mathrm{b}$, the baryon density at the current time,
    \item [ii]  $\Omega_\mathrm{cb}$, the total matter density excluding  massive neutrinos at the current time,
    \item [iii]  $H_\mathrm{0}$, the Hubble parameter at the current time,
    \item [iv]   $n_{s}$, the spectral index of the primordial power spectrum,
    \item [v]    $A_{s}$, the amplitude of the primordial power spectrum,
    \item [vi]   $w_{0}$, the value of the equation of state at the current time,
    \item [vii]  $w_{a}$, the time-dependent part of dark energy,
    \item [viii] $\sum m_{\nu}$, the sum of masses of all neutrinos.
\end{itemize}

We assume a flat Universe:
\begin{align}
\label{eq:flat}
\Omega_\mathrm{rad} + \Omega_\mathrm{m} + \Omega_\mathrm{DE} &= 1\ ,
\end{align}
where $\Omega_\mathrm{rad}$, $\Omega_\mathrm{m}$, and $\Omega_\mathrm{DE}$ represent the total radiation density including the relativistic neutrinos, the total matter density including non-relativistic neutrinos, and the dark energy density, respectively.
We set $T_\mathrm{CMB} = 2.7255\mathrm{K}$ to obtain the radiation density,
\begin{equation}
\Omega_\mathrm{rad} = (1 + \frac{7}{8} N_\mathrm{ur} \Gamma_{\nu}^{4}) \Omega_{\gamma}\ ,
\end{equation}
where we set $N_\mathrm{ur} = 2.0328$ for most cosmologies with a single massive neutrino species. 
$\Gamma_{\nu} = (4/11)^{1/3}$ represents the neutrino-to-photon temperature ratio today.
The total matter density includes two non-relativistic components, $\Omega_\mathrm{m} = \Omega_\mathrm{cb} + \Omega_{\nu}$, where the exact treatment of massive neutrinos in the background is given by
\begin{equation}
\Omega_\nu(a) E^2(a)=\frac{15}{\pi^4} \Gamma_\nu^4 \frac{\Omega_{\gamma, 0}}{a^4} \sum_{j=1}^{N_\nu} \mathcal{F}\left(\frac{m_j a}{k_B T_{\nu, 0}}\right)\ .
\end{equation}
Here, $E(a) = H(a)/H_\mathrm{0}$ is the normalized Hubble parameter, and $m_j$ is the mass of the neutrino species $j$.
$\mathcal{F}$ is an integral resulting from the Fermi-Dirac distribution \cite{2017MNRAS.466.3244Z}:
\begin{equation}
\mathcal{F}(y) \equiv \int_0^{\infty} d x \frac{x^2 \sqrt{x^2+y^2}}{1+e^x}\ .
\label{eq:femi-dirac}
\end{equation}
For the dynamical dark energy model, we utilize the CPL parametrization $w(a) = w_{0} + w_{a}(1 - a)$, because of its capability to distinguish among diverse dynamical scalar field models for dark energy (e.g., \cite{2001IJMPD..10..213C,2003PhRvL..90i1301L}).
Thus, we decide to design the emulator under this $w_{0}w_{a}\mathrm{CDM} + \sum m_\nu$ model.

\subsection{Parameter Ranges}
\label{sec:parameter_range}

The determination of the cosmological parameter space presents a complex challenge, as it must strike a balance between accommodating a broad range of applications and ensuring the emulator's accuracy. 
With the increasing galaxy number density and the corresponding reduction in statistical noise in contemporary cosmological analyses, several tensions have emerged, highlighting discrepancies with the standard cosmological model (e.g., \cite{2022PhRvD.105b3514A,2022NewAR..9501659P,2022JCAP...02..007W}).
To address these tensions, incorporating constraints from diverse cosmic probes and datasets proves advantageous. 
Thus, the emulator is required to cover a sufficiently broad cosmological parameter space to be applicable to the probes with low statistical significance.
However, exploring a broad cosmological parameter space poses practical challenges,
as it necessitates a large number of training samples to achieve the desired level of accuracy. 
In our case, the number of training simulations is fixed according to the available computational resources. 
To determine the ranges of parameters for the demanded $1\%$ accuracy prediction, we need to construct a surrogate model based on the nonlinear matter power spectrum generated by the HaloFit model in CLASS \cite{2011arXiv1104.2932L,2012ApJ...761..152T}.

The eight cosmological parameters we select here are similar to the settings of \texttt{EuclidEmulator2}, and they can be directly converted into input parameters in numerical simulations. 
For the five standard $\Lambda$CDM parameters such as $\Omega_\mathrm{b},\ \Omega_\mathrm{m}, \ H_\mathrm{0}, \ A_{s}$, and $n_{s}$, the current CMB data provide very tight constraints on them, while the constraints from LSS data are relatively weak. 
To reduce the influence of the prior range during MCMC analysis, we initially selected a parameter range larger than the Planck 2018 constraint result ($\sim 10 \sigma$). 
For the two parameters $w_{0}$ and $w_{a}$ describing dynamical dark energy, we initially selected a range of $4\sigma$ based on the joint constraint result of Planck+SNe+BAO (Table~6 in Planck18 paper \cite{2020A&A...641A...6P}).
Due to the accuracy limitation in treating neutrinos in the simulations (detailed in Section \ref{sec:simu}), the range of the last parameter, the total mass of massive neutrinos, is fixed at [0.00, 0.30] eV.

Then, we selected 128 training samples using Sobol sequence sampling \cite{sobol1967distribution} within the above parameter space.
The Latin hypercube sampling \cite{mckayComparisonThreeMethods1979,tangOrthogonalArrayBasedLatin1993} method is utilized to generate extra validation samples.
We have confirmed that the 68th percentile error of 100 validation samples converged with the result of 1000 validation samples.
Thus, we use 100 validation samples for the determination of parameter ranges.
For each cosmology, we use CLASS\footnote{\url{https://github.com/lesgourg/class_public}} to generate the corresponding linear power spectrum and HaloFit nonlinear power spectrum \cite{2011arXiv1104.2932L}.
The output redshifts are the same as the simulation snapshots detailed in Section~\ref{sec:nbody}, and $3072$ points are taken at equal intervals in $k\in [0.006, 19.3]\ \hmpc$.
Combining PCA and GPR, we emulate the ratio of the nonlinear power spectrum and the linear power spectrum $B(k,z) = P_\mathrm{HaloFit}/P_\mathrm{linear}(k, z)$. 
By adjusting each cosmological parameter's range and checking whether the result meets the requirement that the 68th percentile error of the test samples is less than $1\%$, we determine the range of eight variables as follows:

\begin{equation}
\label{eq:parameter-space}
\begin{aligned}
\Omega_{\mathrm{b}} & \in[0.04, 0.06]\ , \\
\Omega_{\mathrm{cb}} & \in[0.24, 0.40]\ , \\
n_{\mathrm{s}} & \in[0.92, 1.00]\ , \\
H_\mathrm{0} & \in[60, 80]\ \mathrm{km\, s^{-1}\, Mpc^{-1}}\ , \\
A_{s} & \in[1.70, 2.50] \times 10^{-9}\ , \\
w_{0} & \in[-1.30, -0.70]\ , \\
w_{a} & \in[-0.50, 0.50]\ , \\
\sum m_{\nu} & \in [0.00, 0.30]\ \mathrm{eV}\ .
\end{aligned}
\end{equation}
Compared with \texttt{EuclidEmulator2}, the ranges of most parameters are similar. 
However, the range of $w_{a}$ is slightly narrowed because when $w_{a} + w_{0}\rightarrow0$, the growth of structure and expansion of the universe deviates significantly from the standard cosmology.
The overall performance of the cosmological emulator will be affected significantly.
$H_\mathrm{0}$ has a relatively larger range to reduce the prior effect of parameters in studies of the Hubble tension.
Note that we extended the parameter range of neutrino mass to twice that of \texttt{EuclidEmulator2}. 
This is due to the recently developed algorithm \cite{2022JCAP...09..068H} used in these simulations, which incorporates the effect of massive neutrinos on pure dark matter simulations.
Percent-level accuracy has been validated for $\sum m_{\nu} \leq 0.3 \mathrm{eV}$ in the Euclid neutrino code comparison paper \cite{2023JCAP...06..035.}. 
We will discuss more details about this method later.

\subsection{Sampling}
\label{sec:sampling}

Generally, we assume that there is no prior knowledge about the dependence of simulated statistics on given parameters, meaning that all parameters follow a uniform distribution within the given interval $[a_i, b_{i}]$.
The next step is to determine an appropriate sampling technique in the normalized parameter space.
This is a critical aspect of emulator design, as the training data's quality and diversity directly impact the final model's accuracy and versatility.
In the earliest \texttt{Coyote} Universe \cite{2010ApJ...715..104H,2009ApJ...705..156H,2010ApJ...713.1322L,2014ApJ...780..111H,2016JCAP...08..008C}, Latin hypercube sampling was used to construct a five-parameter emulator.
Many other emulators also use this method for experimental design with different modifications (e.g., \texttt{Aemulus} \cite{2019ApJ...875...69D},
\texttt{Dark Quest} \cite{2019ApJ...884...29N},
\texttt{BACCO} \cite{2021MNRAS.507.5869A},
\texttt{EuclidEmulator} \cite{2019MNRAS.484.5509E,2021MNRAS.505.2840E}).
We refer interested readers to the recent review \cite{moriwakiMachineLearningObservational2023} for a more detailed comparison.

However, both the parameter dimension and number of samples must be determined at the beginning of the Latin hypercube sampling.
This means that it is non-trivial to directly insert more sample points to increase the emulator accuracy, in case the accuracy requirement is increased in the near future.
A natural extension to the established cosmological parameter space is to emulate galaxy clustering with the halo occupation distribution (HOD \cite{2005ApJ...633..791Z}) model or to incorporate baryonic effects with various correction methods (e.g., \cite{2015JCAP...12..049S,2019JCAP...03..020S,2021MNRAS.506.4070A,2021JCAP...12..046G,2023MNRAS.523.2247S}).
In this case, the sequence sampling methods, commonly used in the quasi-Monte Carlo problems, attract our attention.
Their deterministic generation process makes sample points independent of the parameter dimension and the number of samples (see \ref{sec:sobol} for an example).
At the same time, their low discrepancy can provide an efficient space-filling strategy in the high-dimensional space.

In this paper, we mainly focus on the Sobol sequence \cite{sobol1967distribution} and the Halton sequence \cite{halton1960efficiency}.
The former utilizes a base of 2 to fill successively the whole unit hypercube and reorders the sequence in each dimension.
In contrast, the latter employs distinct prime bases per dimension but suffers from correlation issues in higher dimensions.
Both methods have been used in cosmological analysis for different purposes.
In 2022, the \textsc{CosmoGridV1} simulation suite generated 2500 cosmologies sampled on a six-dimensional Sobol sequence for map-level cosmological inference \cite{2023JCAP...02..050K}.
\texttt{Aemulus-$\nu$} simulation suite employed a two-tiered experimental design, with the second tier simulations sampled using a Sobol sequence \cite{2023JCAP...07..054D}.  
Recently, the \textsc{Quijote} project \cite{2020ApJS..250....2V} also provided a Sobol sequence set, including $32,768$ $N$-body simulations designed for machine learning applications\footnote{\url{https://quijote-simulations.readthedocs.io/en/latest/bsq.html}}.
The Halton sequence, on the other hand, is utilized in the generation of random samples for correlation function calculation to improve the accuracy \cite{2022A&A...666A.181K}.

\begin{figure*}[!htbp]
    \centering
    \includegraphics[width=0.95\textwidth]{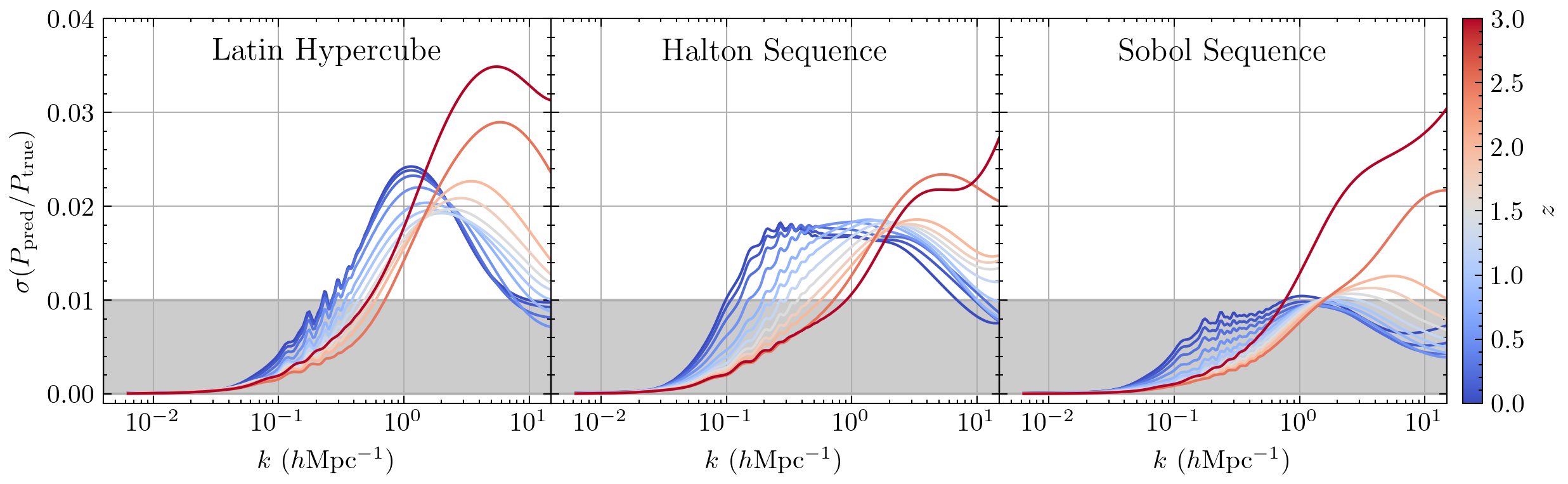}
    \caption{The 68th percentile fractional errors for three different sampling schemes.
    The number of training samples is fixed at 128.
    Different colors represent different redshifts.
    Sobol sequence sampling has the best performance.
    }
    \label{fig:latin_halton_sobol}
\end{figure*}

To test the impact of different sampling methods on the final prediction accuracy, we utilized Latin hypercube, Sobol sequence, and Halton sequence sampling to generate 128 training cosmologies, respectively.
The construction and validation processes follow the same procedure as the determination of parameter ranges in Section \ref{sec:parameter_range}.
The results are shown in Fig.~\ref{fig:latin_halton_sobol}.
The emulation accuracy of different redshifts is represented by the 68th percentile fractional errors of validation samples.
We find the utilization of the Sobol sequence significantly improves the prediction accuracy of the power spectrum at $z\leq 2$, compared with the other two methods.
The number of training samples was varied for all three different methods.
A better convergence rate is also observed for the Sobol sequence sampling (not shown here).

\begin{figure*}[!htbp]
    \centering
    \includegraphics[width=0.9\textwidth]{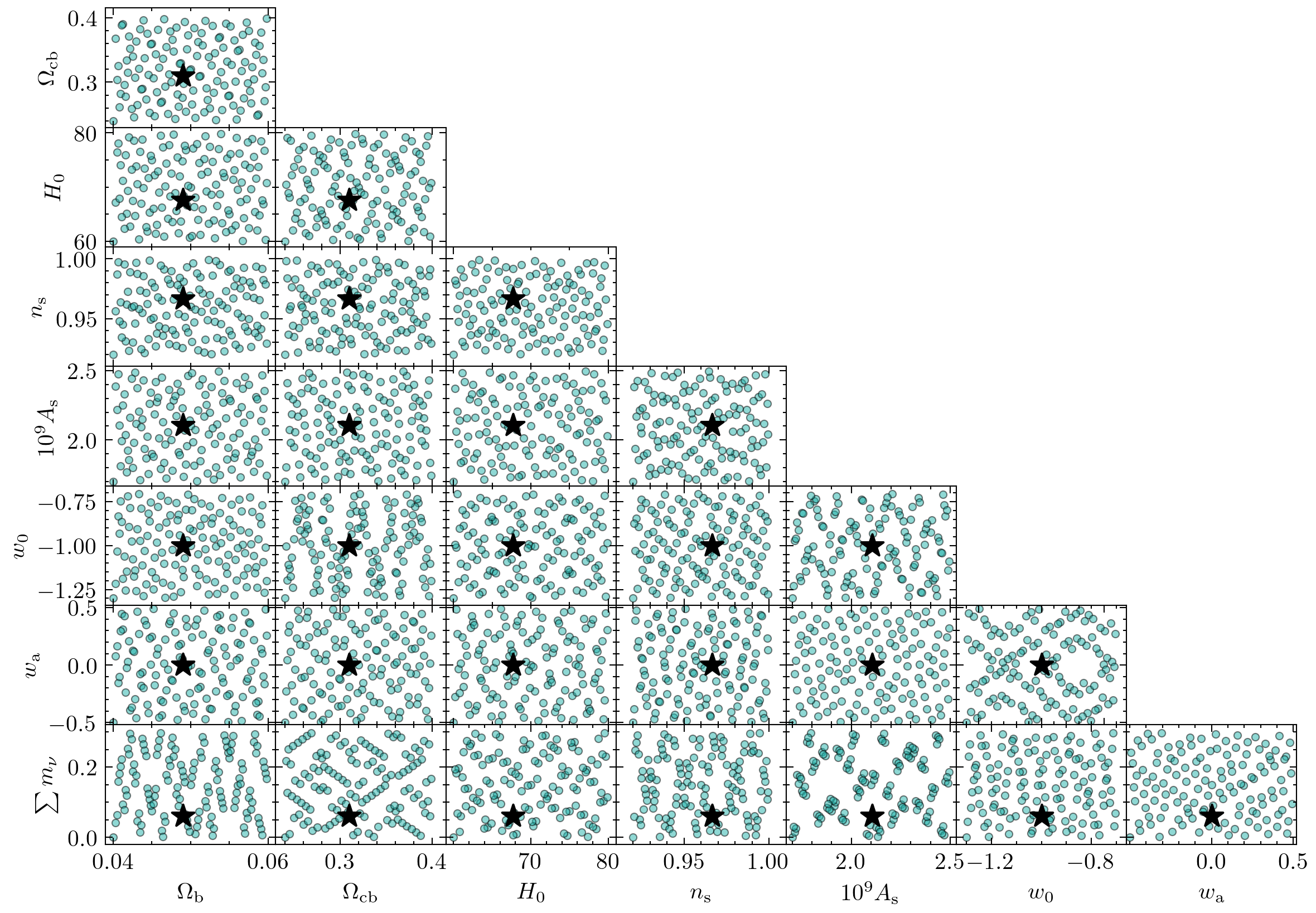}
    \caption{129 cosmologies used in \texttt{CSST Emulator}. The black star represents the fiducial c0000 cosmology. The Sobol sequence samples 128 other cosmologies, which are shown by dots. The special patterns in some planes arise from the projection effect.}
    \label{fig:cosmologies}
\end{figure*}

Therefore, we sampled 128 grid points using an eight-dimensional Sobol sequence within the previously defined cosmological space.
An extra simulation is generated as our fiducial model.
The cosmology is selected from the last column in Table~2 in the Planck 2018 results \cite{2020A&A...641A...6P}.
All cosmologies in our emulator are shown in Fig.~\ref{fig:cosmologies}, with
the fiducial Planck 2018 cosmology (c0000) and other Sobol sequence cosmologies (c0001-c0128) represented by a star and dots, respectively.
The special patterns in some planes (e.g., $A_{s}\mbox{-}\sum m_{\nu}$ plane) arise from the projection effect, and the apparent clustered points are distant in the high-dimensional space.

\section{Cosmological Simulation Suite}
\label{sec:simu}

$N$-body simulations generate the training data for cosmological emulators, which is the most important and computationally intensive step.
The power of precise nonlinear prediction is directly derived from the simulation data.
In this section, we describe the \textsc{Kun} cosmological simulation suite, which provides the training data used in our emulator.

\subsection{$N$-body Simulations}
\label{sec:nbody}

In cosmological numerical simulations, two critical parameters are the simulation volume and the particle count. 
For simulations with a fixed cubic side length $L$, cosmological information at wave numbers $k < 2\pi/L $ is missing. 
The absence of long-wavelength fluctuations in a finite simulation volume results in systematic inaccuracies in both the power spectrum and its covariance. 
Various studies have demonstrated that a simulated volume of $V \gtrsim 1\,\gpcht $ and a mass resolution of approximately $M_\mathrm{p} \sim 10^9\,h^{-1} \Msun $ are necessary for Stage-\Rmnum{4} surveys to achieve convergence at the percent level for various measurements and statistical uncertainties (e.g., \cite{2016JCAP...04..047S,2019MNRAS.489.1684K}).
Therefore, we run each simulation with $3072^3$ particles in a $1\,\gpcht$ box, indicating a particle mass of $2.87 \frac{\Omega_\mathrm{cb}}{0.3} \times 10^{9}\, h^{-1}\Msun$.
This enables our simulations to obtain precise predictions at the extremely nonlinear region ($k \lesssim 10\,\hmpc$).
For each cosmology, we generate only a single simulation.
In order to suppress the cosmic variance at large scales, the fixed amplitude method \cite{2016MNRAS.462L...1A} is employed to set up the initial density field.

We utilize the Gadget-4\footnote{\url{https://wwwmpa.mpa-garching.mpg.de/gadget4/}} $N$-body solver to run all numerical simulations \cite{2021MNRAS.506.2871S} and modify the background evolution part so that the code can calculate the expansion rate $H(a)$ and the growth factor $D_{+}(a)$ under the $w_{0}w_{a}\mathrm{CDM} + \sum m_\nu$ cosmology exactly\footnote{\url{https://github.com/czymh/C-Gadget4}}.
The initial condition generator is also modified correspondingly.
The recent Newtonian motion gauge is utilized to incorporate the massive neutrino effect on the cold matter particles in our training simulations \cite{2020JCAP...09..018P,2022JCAP...09..068H}. 
We will discuss more details in Section \ref{sec:neutrino}.

All particles are initialized with the second-order Lagrangian Perturbation Theory (2LPT \cite{1998MNRAS.299.1097S,2006MNRAS.373..369C}) at the fixed redshift $z = 127$.
The convergence test on the setup of initial redshift is detailed in Section \ref{sec:convergence_test}.
An isotropic glass-like distribution is applied to pre-initial loads.
At small scales, we fix the equivalent Plummer softening length to $\epsilon = 8~\kpch$ in comoving coordinates.
At large scales, the grid size for Particle-Mesh (PM) force calculation is fixed at $N_\mathrm{grid}=6144$.
The maximum allowed time step is set to $\mathrm{max}(\Delta \ln a) = 0.04$ and we verified that the matter power spectrum in both real space and redshift space converges to the results obtained with $\mathrm{max}(\Delta \ln a) = 0.01$ at $z\leq 3$ and $k \leq 10\,\hmpc$.

We output snapshots at 12 fixed redshifts of $z = $\{3.00, 2.50, 2.00, 1.75, 1.50, 1.25, 1.00, 0.80, 0.50, 0.25, 0.10, 0.00\}.
Both \textsc{SubFind} \cite{2001MNRAS.328..726S} and \textsc{Rockstar} \cite{2013ApJ...762..109B} are utilized on each time snapshot for dark matter halo and subhalo identifications.

We also save two on-the-fly pyramid particle light-cones up to $z\leq 3$ by tiling simulation boxes along the three main axes for future weak lensing research.
Each cone covers $2116\,\mathrm{deg^2}$.
To dilute the structure repetitions caused by box replication, we select two special line-of-sight directions $\vec{n}_1 = \{ 0.8678,\ 0.3595,\ 0.3431 \}$ and $\vec{n}_2 = \{ 0.3595,\ 0.8678,\ 0.3431\}$ \cite{2024MNRAS.534.1205C}.
Besides the two particle light-cones, the full-sky projected density maps are preserved in a HEALPix scheme with a thickness of $50\,\hmpc$ at the same redshift range.
The corresponding angular resolution is $\Delta \theta= 0.43\,\mathrm{arcmin}$ with $N_\mathrm{side} = 8192$.
The whole simulation set costs about $1.5\times 10^{8}$ CPU hours and the storage is about $3.1$ PB.
This \textsc{Kun} simulation suite will be made publicly available according to reasonable requests.

\subsection{Neutrinos}
\label{sec:neutrino}

The inclusion of neutrinos affects both the expansion rate and structure growth of the Universe.
Newtonian motion gauges enable us to capture the full impact of linear neutrino perturbations on the nonlinear evolution of cb particles in an ordinary Newtonian $N$-body simulation \cite{2020JCAP...09..018P,2022JCAP...09..068H}.
Results for several statistics of both matter and haloes have demonstrated adequate accuracy compared to particle-based and mesh-based methods \cite{2023JCAP...06..035.}.
The basic idea is to incorporate the additional potential contribution of neutrinos via a gauge transformation.
We can achieve this by using the three steps detailed below.
\begin{itemize}
    \item [  i)] Modify the Hubble rates $H(a)$ to the exact form, including radiation component and relativistic neutrinos in the $N$-body solver:
    \begin{equation}
    \label{eq:Friedman}
        E^2(a) 
        = \Omega_{\mathrm{cb}} a^{-3}+\Omega_\nu(a)+\Omega_{\mathrm{rad}} a^{-4}+\Omega_{\Lambda}\ .
    \end{equation}
    
    \item [ ii)] Backscaling the linear cb power spectrum generated by CLASS\footnote{\url{http://class-code.net/class.html}} at the late time to the initial redshift $z=127$ with the scale-independent growth factor $D_{+}$ from Eq.~\ref{eq:Friedman}. 
    Note that we set the late-time reference redshift $z_\mathrm{low} = 1.0$.
    Then we evolve the initial particles using the Newtonian solver.
    
    \item [iii)] The output is saved in the coordinates of the Newtonian motion gauge.
    The particle positions can be transformed to the usual N-boisson gauge according to a displacement field calculated from the Einstein–Boltzmann solver outputs (detailed in Appendix A of \cite{2022JCAP...09..068H}).
\end{itemize}

The effect of linear neutrino perturbation is by construction fully included at the reference redshift $z_\mathrm{low}=1.0$.
For other redshifts, a coordinate transformation is required to incorporate the residual impact of neutrinos.
This small correction primarily affects the power on large scales while becoming negligible on small scales, which are of greater interest in this work.
Therefore, we omit the final post-processing step in the entire simulation suite (also done in \cite{2023JCAP...06..035.}).
In this situation, $z_\mathrm{low}=1.0$ is a reasonable choice to ensure that the correction is small for both the highest and lowest redshifts ($z=3.0$ and $z=0.0$ in our case. See Fig.~6 in \cite{2022JCAP...09..068H} for detailed comparisons of different $z_\mathrm{low}$ values).

\subsection{Convergence Tests}
\label{sec:convergence_test}

The emulator can only provide reliable predictions when the training simulations have converged statistics at the redshifts and scales of interest.
The accurate initial condition is essential for a cosmological simulation.
There are two nuisance parameters: 1) the given order of LPT, and 2) the initial redshift to start evolving all particles in this process.
We chose to initialize all simulations with 2LPT for the first parameter.
For the second parameter, we run three extra test simulations to show the convergence of our fiducial setup.
All the test simulations adopt the Planck 2018 cosmology without massive neutrinos \cite{2020A&A...641A...6P}. 
We maintain the particle mass resolution $m_\mathrm{part} = 2.98 \times 10^{9}\, h^{-1}\Msun$ consistent with the fiducial setup but with a smaller volume $(250\,\mpch)^{3}$ to save the computational costs.
The cosmic variance is not important in this convergence test as all the test simulations use the same initial Gaussian random field.
The fiducial initial redshift is fixed at 127 for the first convergence test simulation, CT0.
The other two simulations start at $z_\mathrm{ini} = 63$ (CT1) and $z_\mathrm{ini} = 255$ (CT2).

\begin{figure*}
    \centering
    \includegraphics[width=0.9\textwidth]{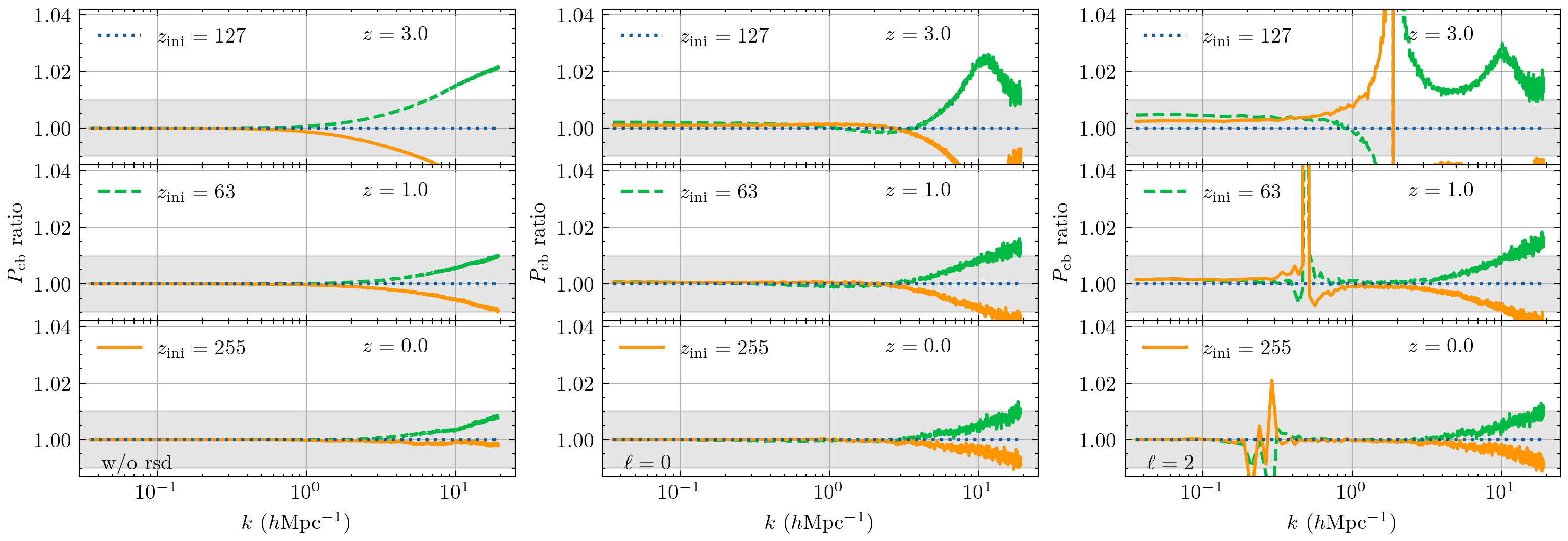}
    \caption{Comparisons of both real-space dark matter power spectra (left) and redshift-space spectra monopole (middle), quadrupole (right) between the CT1 simulation initialized at $z_\mathrm{ini}=63$ and the CT2 simulation initialized at $z_\mathrm{ini}=255$ to our fiducial CT0 simulation started at $z_\mathrm{ini}=127$. All simulations are initialized with 2LPT. The grey shadow indicates the $1\%$ difference. 
    The jump in the quadrupole ratio is due to the zero-crossing and is not important.}
    \label{fig:pk-zini}
\end{figure*}

We compare cb power spectra measured from CT1 and CT2 simulations with the CT0 results in Fig.~\ref{fig:pk-zini}.
A comparison of power spectra in real space is shown in the left panel, with different rows representing different redshifts.
We find the $1\%$ convergence (grey region) for $k \leq 10\, \hmpc$ and $z < 3.0$ across the three test simulations.
The difference becomes larger for higher redshift, primarily due to the shorter time for the discreteness effect/decaying mode to vanish.
For the highest redshift $z=3$, the difference slightly exceeds $1\%$ at $k\sim 10 \,\hmpc$ where shot noise becomes dominant.
The redshift-space cb power spectra monopole and quadrupole are shown in the middle and right panels, respectively.
A consistent convergence level is observed.
In conclusion, starting the simulation at $z_\mathrm{ini} = 127$ with 2LPT is a reasonable choice for our purpose.

\section{Emulator Construction}
\label{sec:emulation}

In this section, we describe how to obtain the smooth matter power spectrum, construct an emulator by combining PCA with GPR, and evaluate the performance without extra validation samples.

\subsection{Smooth Spectra}
\label{sec:smooth}

In this work, we calculate the power spectrum using Nbodykit\footnote{\url{https://github.com/bccp/nbodykit}} \cite{2018AJ....156..160H}, with $k_\mathrm{min} = 0.0063\,\hmpc, \ k_\mathrm{max} = 10\,\hmpc$ and $\mathrm{d}k = 0.003\,\hmpc$, resulting in 3182 data points for each snapshot.
The mesh size is fixed at $N_\mathrm{mesh}=3072$.
The Cloud-In-Cell (CIC) mass assignment scheme is used, and the compensation is applied in the calculation.
Besides, we employ the interlacing technique to reduce the aliasing effect \cite{hockney2021computer}.
As a basic test, the measured spectra with $N_\mathrm{mesh}=6144$ under the fiducial cosmology are utilized to compare with our fiducial results.
We find the maximum error at $k=10\,\hmpc$ is within 0.5\%, indicating that our choice in the power spectrum analysis is reasonable.
However, more caution should be taken for the samples with low number density or with the redshift space distortions (e.g., \cite{2024JCAP...09..044W}).

Although the fixed amplitude technique significantly suppresses cosmic variance, the residual remains substantial, exceeding $1\%$ at the BAO scale.
To further reduce cosmic variance, paired simulations could be employed, but this would double the computational resource consumption.
To avoid this, we utilize the matched and paired fast simulations to mitigate fluctuations at large scales.

For each simulation, we generate two extra low-resolution simulations by utilizing the efficient PM code FastPM \cite{2016MNRAS.463.2273F}.
The first one (FA1) uses the same initial Gaussian field as the high-resolution $N$-body simulation but discards high-frequency modes.
The second one (FA2) is initialized with the inverted-phase Gaussian field by converting $\theta \rightarrow \pi + \theta$ \cite{2016PhRvD..93j3519P}.
We can obtain the transfer function at each redshift from the matched pair:
\begin{equation}
\label{eq:transfer_fa}
T(k,z) = P_\mathrm{Gadget4}(k, z)/P_\mathrm{FA1} (k, z)\ .
\end{equation}
Then the paired power spectrum can be measured by:
\begin{equation}
\label{eq:paired}
P_\mathrm{paired} (k,z) = T(k,z)\times \left(P_\mathrm{FA1} (k, z) + P_\mathrm{FA2} (k, z) \right)/2\ .
\end{equation}
In principle, the transfer function can compensate for the loss of power at intermediate to small scales in FastPM simulations and does not suffer any cosmic variance.
Thus, the paired results should be the same as the original fixed \& paired results in \cite{2016MNRAS.462L...1A}.

The main purpose of the FastPM runs is to beat down the cosmic variance at large scales.
We have confirmed that a $768^3$ particle load is sufficient to obtain a reliable power spectrum at $k \leq 0.2\,\hmpc$ across all redshifts.
The force resolution factor ($B$), the number of time steps ($N_{\rm step}$), and the initial redshift are varied to identify the optimal values for better convergence with the full result.
We find that starting the fast simulations at $z=19$ with $B=2$ and $N_{\rm step} = 40$ time steps linearly spaced in scaling factor $a$ is the best choice for our purpose.

\begin{figure}[H]
    \centering
    \includegraphics[width=0.4\textwidth]{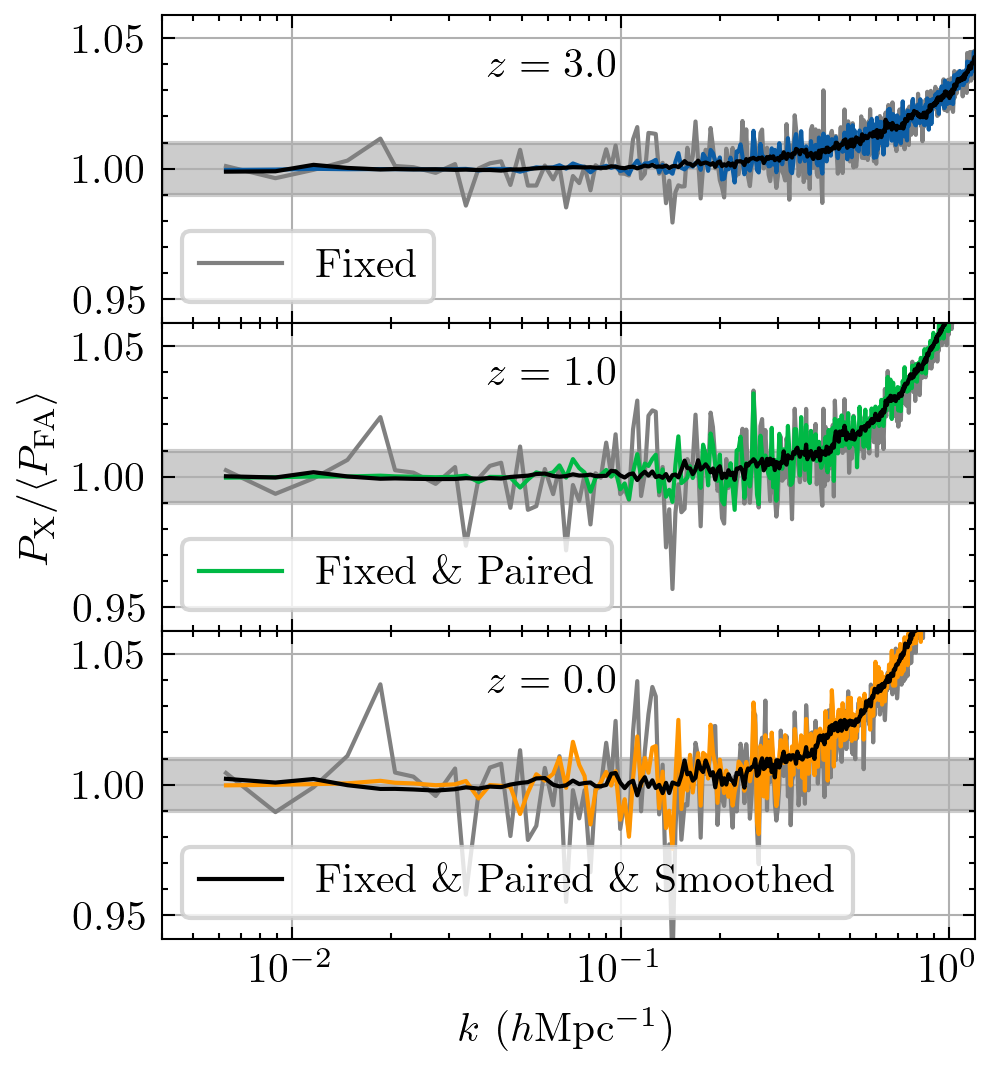}
    \caption{ The fixed, paired, and smoothed matter power spectra normalized by the averaged power of 25 pairs of FastPM simulations.
    The grey region indicates the $1\%$ difference.
    Paired power spectra (colored lines) have much less cosmic variance fluctuations than fixed-only spectra (grey lines), especially at BAO scales.
    Savitsky-Golay smoothing further reduces the fluctuations at all scales (black lines).}
    \label{fig:pk-smooth}
\end{figure}

To evaluate the performance of our variance reduction procedure, we generate an extra 25 pairs of FastPM simulations at the fiducial c0000 cosmology with fixed \& paired initial conditions.
Each pair runs with different random seeds.
The colored lines in Fig.~\ref{fig:pk-smooth} represent the ratio between the paired spectra obtained from two extra FastPM runs (Eq.~\ref{eq:paired}) and the mean spectra of all 50 fast simulations.
The grey lines on behalf of the fixed-only results are also shown for comparison.
We find the fixed \& paired technique can effectively reduce most of the cosmic variance, particularly at high redshifts.

However, there is still roughly $1\%$ noise for the spectra at $z=0.0$.
Similar as \cite{2023JCAP...07..054D}, a third-order Savitsky-Golay filter\footnote{\url{https://docs.scipy.org/doc/scipy/reference/generated/scipy.signal.savgol_filter.html}} \cite{1964AnaCh..36.1627S} with a window length of 33 (i.e., the number of nearby data points) is employed to the ratio of paired spectra to the 1-loop LPT power spectra from \textsc{velocileptors}\footnote{\url{https://github.com/sfschen/velocileptors}} \cite{2020JCAP...07..062C,2021JCAP...03..100C} to remove residual fluctuations.
The smoothed results are shown by black solid lines in Fig.~\ref{fig:pk-smooth}.
We have tested that this smoothing process does not introduce additional systematic biases across all cosmologies.

\subsection{Emulation}
\label{sec:emupk}

After measuring the smoothed matter power spectra for each cosmology, we can construct our final emulator now.
At first, finding a better representation of the power spectrum is expected to enhance the performance of emulation (e.g., \cite{2009ApJ...705..156H}).
For instance, the \texttt{CosmicEmu}, including \texttt{Coyote} and \texttt{Mira-Titan} Universe, emulates the rescaled power spectrum $k^{3/2} P(k)$ in logarithmic space to better predict baryonic acoustic oscillations.
For the \texttt{EuclidEmulator}, the nonlinear power spectrum is divided by its linear counterpart, $B(k, z) = P_\mathrm{simu}(k,z)/P_\mathrm{linear}(k,z)$.
The \texttt{BACCO} emulator and \texttt{Aemulus-$\nu$} emulator replace the linear power spectrum in the denominator with the BAO smeared linear and 1-loop LPT power spectrum, respectively.
A common understanding is that the ratio quantity is easier to emulate because separating training statistics into well-known parts and less-understood components significantly reduces the complexity of the target function.

\begin{figure}[H]
    \centering
    \includegraphics[width=0.4\textwidth]{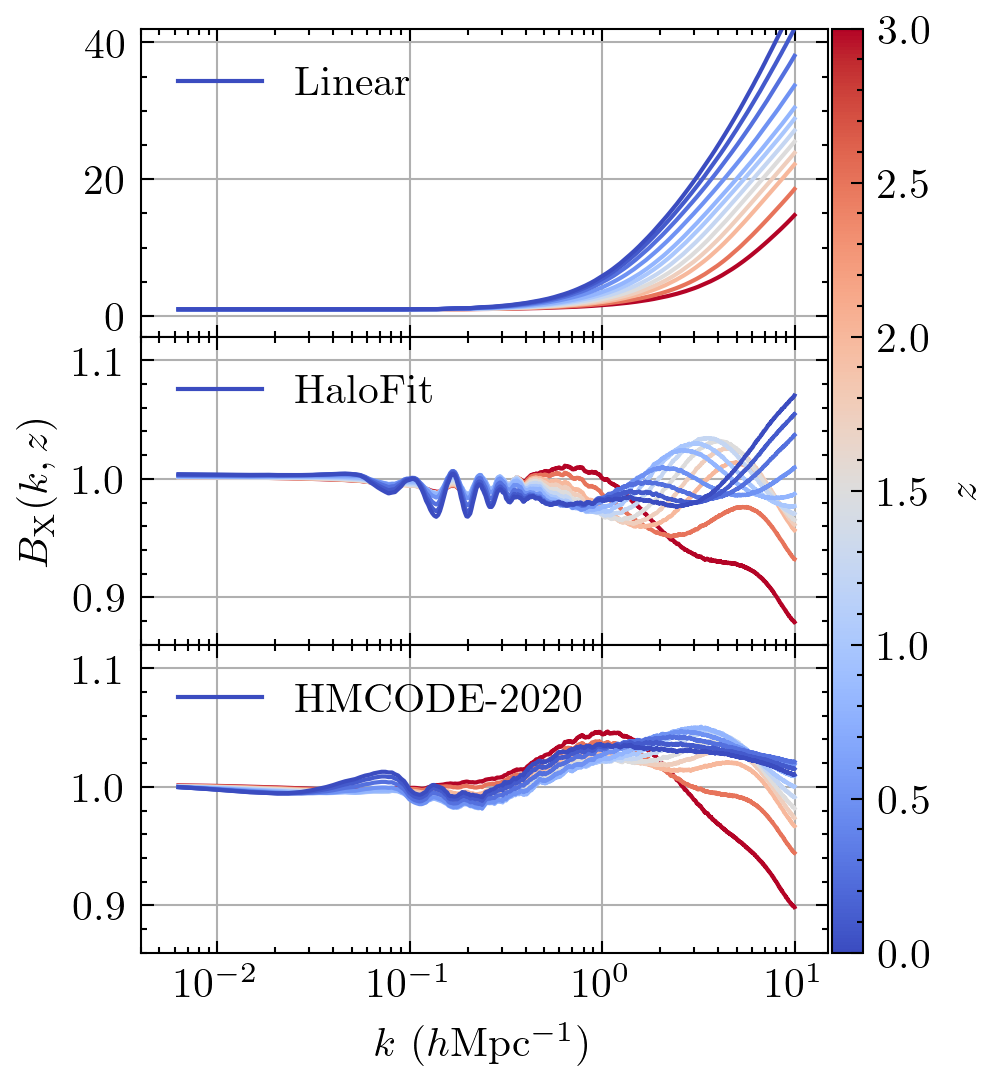}
    \caption{The ratios of the simulated power spectrum to linear (top), HaloFit (middle), and HMCODE-2020 (bottom) predictions under the c0000 cosmology. The color map indicates the redshift in the range of $z\in [0, 3]$.}
    \label{fig:B_X}
\end{figure}

In this paper, we try three different choices of the denominator: linear power spectrum, the nonlinear prediction from revised HaloFit (Takahashi et al.~2012 \cite{2012ApJ...761..152T}), and HMCODE-2020 (Mead et al.~2020 \cite{2021MNRAS.502.1401M}), defined as
\begin{equation}
\label{eq:Bk_X}
B_\mathrm{X}(k,z) = \frac{P_\mathrm{simu} (k,z)}{P_\mathrm{X}(k,z)}\ ,
\end{equation}
where $\mathrm{X} = \{ \mathrm{linear,\ HaloFit,\ \text{HMCODE-2020}}  \}$.
The ratios in our fiducial cosmology are illustrated in Fig.~\ref{fig:B_X}.
Compared with the linear prediction as the denominator, the HaloFit and HMCODE-2020 results are much closer to unity, indicating more cosmological information captured by the nonlinear models.

In total, we obtain $ N_\mathrm{snap} \times N_\mathrm{vec} $ emulated points, where $N_\mathrm{snap} = 12$ is the number of snapshots per simulation, and $N_\mathrm{vec} = 3182$ is the number of wavenumbers.
PCA technology is then applied to reduce the cross-correlation and dimensionality of the training set.
All the data is decomposed as
\begin{equation}
\label{eq:pca}
B_\mathrm{X}(k,z;\hat{\theta}) = \mu_{B_\mathrm{X}}(k,z) + \sum_{i=1}^{N_\mathrm{PCA}} \phi_{i}(k,z) w_{i}(\hat{\theta})\ , 
\end{equation}
where $\hat{\theta} \in [0,1 ]^{n_{\hat{\theta}}}$ is the normalized cosmological parameter space with $n_{\hat{\theta}}=8$.
The mean of emulated data is given by $\mu_{B_\mathrm{X}}(k,z)$.
$w(\hat{\theta})$ represents the coefficients with the shape of $(N_\mathrm{sim}, N_\mathrm{PCA})$ capturing the cosmology dependence of the training data,
and $\phi(k,z)$ represents the basis functionals calculated from singular value decomposition.
In this work, we take the first $N_\mathrm{PCA} = 20$ principal components, which can reconstruct $B_\mathrm{X}(k,z;\hat{\theta})$ with approximately $0.5\%$ error.
This is an adequate choice for our emulation process.

The final high-dimensional interpolation is performed on the coefficients $w(\hat{\theta})$.
We employ the supervised learning method GPR (e.g., \cite{williams2006gaussian}) to each coefficient vector in this step.
This method assumes that the input function follows the multidimensional Gaussian distribution.
\begin{equation}
\label{eq:gaussian_function}
f(\hat{\theta}) \sim \mathcal{G P}\left(m(\hat{\theta}), k\left(\hat{\theta}, \hat{\theta}^{\prime}\right)\right)\ ,
\end{equation}
where $m(\hat{\theta})$ is the mean function, and $k\left(\hat{\theta}, \hat{\theta}^{\prime}\right)$ denotes the covariance function of data points at $\hat{\theta}$ and $\hat{\theta}^\prime$.
In practice, we assume a regression mode with Gaussian noise: $w(\hat{\theta}) = f(\hat{\theta}) + \epsilon$, where $\epsilon \sim \mathcal{N}\left(0, \sigma_n^2\right)$.
To increase the numerical stability, we normalize the input data before the inference, indicating $m(\hat{\theta}) = 0$.
Thus, we obtain the joint Gaussian distribution of the emulated data and the predicted values $f_{*}$ at the test cosmology $\hat{\theta}_{*}$,
\begin{equation}
\left[\begin{array}{c}
w \\
f_{*}
\end{array}\right] \sim \mathcal{N}\left(0,\left[\begin{array}{cc}
k(\hat{\theta}, \hat{\theta})+\sigma_n^2 I & k\left(\hat{\theta}, \hat{\theta}_*\right) \\
k\left(\hat{\theta}_*, \hat{\theta}\right) & k\left(\hat{\theta}_*, \hat{\theta}_*\right)
\end{array}\right]\right)\ .
\end{equation}
Then we can express the conditional distribution of the key predictive functions for this regression as follows.
\begin{equation}
{f}_* \mid \hat{\theta}, {w(\hat{\theta})}, \hat{\theta}_* \sim \mathcal{N}\left(\bar{{f}}_*, \operatorname{Cov}\left({f}_*\right)\right)\ ,
\end{equation}
where
\begin{equation}
\bar{f}_*={k(\hat{\theta}, \hat{\theta}_*)}^{\top}\left[k(\hat{\theta},\hat{\theta})+\sigma_n^2 I\right]^{-1} {w(\hat{\theta})}\ ,
\end{equation}
\begin{equation}
\text{Cov}\left[f_*\right]=k\left({\hat{\theta}}_*, {\hat{\theta}}_*\right)-{k(\hat{\theta}_*, \hat{\theta})}\left[k(\hat{\theta},\hat{\theta})+\sigma_n^2 I\right]^{-1} {k(\hat{\theta}, \hat{\theta}_*)}\ .
\end{equation}

The remaining task is to determine the covariance matrices of training and validation sets by selecting an appropriate kernel function.
In the general case, the correlation between the test position and training points is strongly related to the distance between them in the parameter space.
Therefore, the most commonly used radial basis function (RBF) kernel can be expressed as
\begin{equation}
\label{eq:RBF}
k_\mathrm{RBF}\left(\hat{\theta}_i, \hat{\theta}_j\right)=\exp \left(-\frac{d\left(\hat{\theta}_i, \hat{\theta}_j\right)^2}{2 l^2}\right)\ .
\end{equation}
Here, $l$ is an eight-dimensional vector that describes the response length scale in this work.
For our purpose, we find the combination of an RBF kernel and a constant kernel, $k = C \cdot k_\mathrm{RBF}\left(\hat{\theta}_i, \hat{\theta}_j\right)$, is sufficiently flexible.
The training process in GPR is to find the optimal hyperparameters for the given training data and kernel function.
This is achieved by the optimization of the log marginal likelihood of the input data via maximizing
\begin{equation}
\label{eq:loglikelihood}
\begin{aligned}
\ln \mathcal{L} & =-\frac{1}{2} w(\hat{\theta})^T\left[k\left(w(\hat{\theta}), w(\hat{\theta})\right)+\sigma_n^2 I\right]^{-1} w(\hat{\theta}) \\
& -\frac{1}{2} \log \left\lvert\, k\left(w(\hat{\theta}), w(\hat{\theta}) \right)+\sigma_n^2 I \right\lvert\,-\frac{n}{2} \log 2 \pi\ .
\end{aligned}
\end{equation}
For the practical implementation of PCA and GPR, we utilize the Python library \textsc{scikit-learn} \cite{scikit-learn} in the training process.

After constructing the pipeline of the emulator, the remaining question is to quantify the accuracy of the whole parameter space.
We first employ the leave-one-out (LOO) cross-validation method before validation using extra simulations. 
Each simulation is utilized once as a validation run, while the remaining simulations form the training set. 
Thus, we can obtain the emulation errors for all training cosmologies.
The LOO error is defined as the 68th percentile error ($1\sigma$) across all samples.

\begin{figure*}[!htbp]
    \centering
    \includegraphics[width=0.9\textwidth]{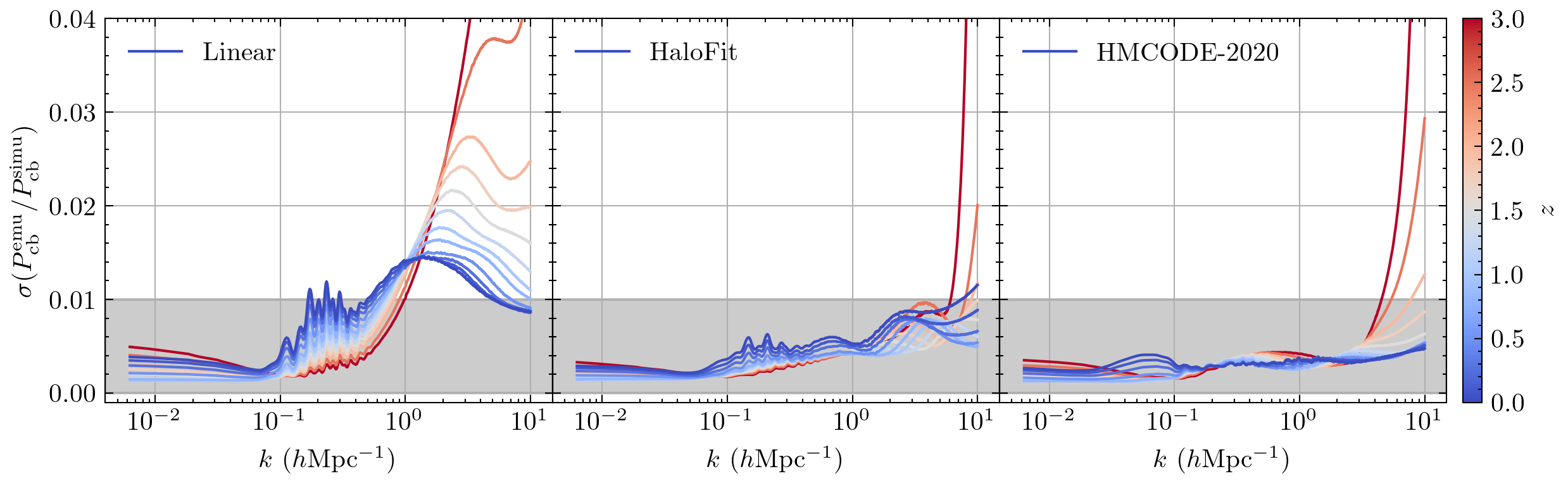}
    \caption{ Leave-one-out errors for three different emulation strategies from $z=0$ (blue) to $z=3.0$ (red).
    The grey shadow indicates $1\%$ errors.
    For the middle and right panel, the results are below the 1\% threshold for $z\leq 2.0$.
    The worse performance at high redshift is likely due to the larger shot noise and smaller power amplitude.
      }
    \label{fig:pkemu-different-methods}
\end{figure*}

The Comparison of performance for cb power spectrum emulations based on three different denominators in Eq.~\ref{eq:Bk_X} is illustrated in Fig.~\ref{fig:pkemu-different-methods}.
The LOO errors given by emulating the ratio to the linear power spectrum are $\leq 2\% $ for all scales at $z\leq 1.5$ (left panel).
The accuracy decreases as redshift increases, which is similar to the trends of \texttt{EuclidEmulator2} results in \cite{2023JCAP...07..054D}.
For the ratio to the HaloFit prediction (middle panel), we can see that it performs better than the previous method.
The HaloFit nonlinear power spectrum is much closer to the simulated power spectrum than linear theory, shown in Fig.~\ref{fig:B_X}.
This halo-model-based function successfully captures complicated cosmological dependence arising from the halo formation physics and is further improved by the calibration using high-resolution simulations.
Thus, the residual information is easier to model with the GPR.
Using the prediction from HMCODE-2020 integrated in the latest CAMB\footnote{\url{https://camb.readthedocs.io/}} code as the denominator works extremely well (right panel).
The error caused by the emulation process is comparable to the simulation error ($\lesssim 1\%$) shown in Fig.~\ref{fig:pk-zini} at $z\leq 2.0$, though slightly worse at higher redshifts.
A possible reason for the higher redshifts is that the shot noises of some cosmologies dominate at small scales, discussed in \ref{sec:shotnoise}.
Finally, we select the HMCODE-based model as our final emulation strategy.
The latter comparison and validation are based on this method.

Note that we provide both HaloFit-based and HMCODE-based emulators in our released code\footnote{\url{https://github.com/czymh/csstemu/}}, as the former performs slightly better than the latter at higher redshifts.
In practice, our emulator integrates all necessary ingredients, enabling it to run without the external packages such as \textsc{scikit-learn} and cosmological Einstein–Boltzmann solvers like CLASS or CAMB,
avoiding 
complexities in package management due to package dependencies.
The denominator is predicted through a similar emulation process, but the number of training cosmologies is increased to $513$ to ensure the $0.5\%$ accuracy for all scales and redshift ranges.
This procedure enables our emulator to provide nonlinear power for a single cosmology within $\lesssim 15\,\mathrm{ms}$ without loss of accuracy, which is $\sim2$ orders of magnitude faster than the Einstein-Boltzmann solvers.
Further details about this code are available at \url{https://csst-emulator.readthedocs.io/}.

\section{Comparison and Validation}
\label{sec:validation}

In this section, we first compare the correlation transformed from the theoretical power spectrum with the results of \texttt{Quijote} \cite{2020ApJS..250....2V} simulations to validate our smoothing procedure detailed in Sec.~\ref{sec:smooth}.
The recent state-of-the-art simulation suite \textsc{FLAMINGO} \cite{2023MNRAS.526.4978S} is utilized to verify our prediction of the spoon-like suppression of massive neutrinos on the power spectrum.
Then the emulator performance is validated by the high-resolution \textsc{CosmicGrowth} \cite{2019SCPMA..6219511J} and \textsc{AbacusSummit} \cite{2021MNRAS.508.4017M} simulation results.
Finally, we compare the accuracy of \texttt{BACCO}, \texttt{EuclidEmulator2}, and \texttt{Mira-Titan IV} (also referred to as \texttt{CosmicEmu}) emulators with the \texttt{CSST Emulator}, using the \textsc{Kun} suite.

\subsection{Correlation Function}
\label{sec:correlation_func}

\begin{figure}[H]
    \centering
    \includegraphics[width=0.4\textwidth]{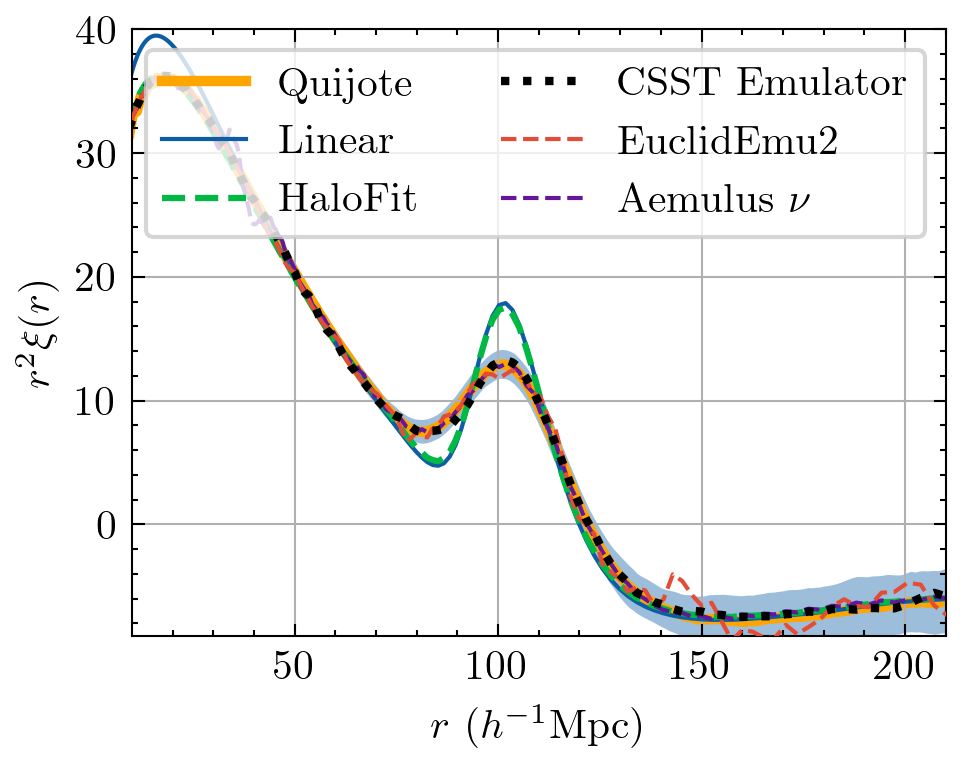}
    \caption{Comparison for matter correlation function at $z=0.0$ of \textsc{Quijote} simulations and other theoretical models.
    The statistical error of 500 fixed and paired \textsc{Quijote} simulations is illustrated by the blue shadow.
    }
    \label{fig:emu_BAO}
\end{figure}

The power spectrum measured from the simulations is affected by cosmic variance on large scales.
To validate the efficiency of our smoothing procedure, we compare the mean correlation function of 250 pairs of fixed \textsc{Quijote} simulations\footnote{\url{https://quijote-simulations.readthedocs.io/}} with the result from our emulator in Fig.~\ref{fig:emu_BAO}.
We present the results only at $z=0.0$ as the sample variance reduction from the paired and fixed amplitude technique \cite{2018ApJ...867..137V} is less efficient at low redshifts.
For a better visualization, the linear prediction is shown by the blue solid line.
The results from HaloFit, \texttt{EuclidEmulator2} and \texttt{Aemulus-$\nu$} are represented by green, red, and purple dashed lines, respectively. 
The correlation functions given by the theoretical tools and emulators are derived from the power spectra by the FFTLog\footnote{\url{https://hankl.readthedocs.io/}} algorithm \cite{1978JCoPh..29...35T,2000MNRAS.312..257H}.

We see that the HaloFit model fails to accurately predict the nonlinear damping of the BAO peak, while the different emulators yield results consistent with the simulations.
The fluctuations at $r\gtrsim 70\,\mpch$ of \texttt{EuclidEmulator2} are attributed to residual noise in the power spectrum.
The excellent agreement between the \texttt{Aemulus-$\nu$} prediction and the \textsc{Quijote} simulations indicates that the Zel'dovich control variates technique removes the majority of sample variance, which is also found in \cite{2023JCAP...07..054D}.
The result given by our emulator exhibits a good convergence with both \textsc{Quijote} and \texttt{Aemulus-$\nu$} results, highlighting the robustness of our smoothing procedure in this work.

\subsection{Neutrino Suppression}
\label{sec:neutrino_suppression}

Massive neutrinos affect both expansion history and structure growth, leading to a suppression of the power spectrum.
Although the suppression also appears for the linear power, nonlinear growth amplifies this effect, reaching a maximum at $k \sim 1\,\hmpc$ in the late Universe.
This results in a spoon-like suppression when comparing the nonlinear power spectrum of cosmology with massive neutrinos and the one with massless neutrinos or smaller mass neutrinos (e.g., \cite{2012MNRAS.420.2551B,2018JCAP...03..049L,2020JCAP...11..062H}).
Current and upcoming Stage-\Rmnum{4} galaxy surveys are expected to measure the sum of neutrino masses and even the neutrino mass hierarchy (e.g., \cite{2022MNRAS.515.5743L,2024arXiv240506047E,2024arXiv240513491E,2024arXiv241102752C}).
Thus, it is crucial to obtain an accurate prediction of neutrino-induced suppression for future surveys.

Since the Newtonian motion gauge only predicts the power spectrum of the $\Omega_\mathrm{cb}$ components,
we adopt the following expression to convert the cb spectra $P_\mathrm{cb}$ into the total matter power spectra $P_\mathrm{mm}$:

\begin{equation}
\begin{aligned}
\label{eq:Pcb2Pmm}
P_\mathrm{mm}(k,z) & = (1-f_\nu)^2 P_\mathrm{cb}(k,z)  + f^2_\nu P^\mathrm{lin}_{\nu\nu}(k,z) \\
& + 2f_\nu(1-f_\nu) \sqrt{P_\mathrm{cb}(k,z)P^\mathrm{lin}_{\nu\nu}(k,z)}\ .
\end{aligned}
\end{equation}

Here, $P^\mathrm{lin}_{\nu\nu}$ denotes the linear prediction of the neutrino auto-power spectrum, obtained from the CLASS.
The neutrino-to-total matter density ratio is defined as $f_\nu = \Omega_\nu / (\Omega_\mathrm{cb} + \Omega_\nu)$.
Since the weak clustering of neutrinos, this approximation yields sufficiently accurate results from large to extremely nonlinear scales.
This approach is also employed by the \texttt{CosmicEmu} emulator and validated by serval papers (e.g., \cite{2008PhRvL.100s1301S,2011MNRAS.410.1647A,2014PhRvD..89j3515U,2015JCAP...07..043C,2016ApJ...820..108H}).
Furthermore, we assume a single massive neutrino component, same as \textsc{AbacusSummit}, while the other emulators commonly assume three neutrino species with equal masses.
We have verified that this assumption introduces only percent-level discrepancies, which are primarily captured by the linear power spectrum.
The nonlinear theoretical tools, such as HaloFit and HMCODE-2020, can predict the power spectra transformation from cosmology with one single massive neutrino to one with three degenerate species with sub-percent accuracy.
Therefore, we calculate this transformation based on the HMCODE-2020 predictions to align our result to the others for comparison\footnote{Here, we do not utilize the HaloFit because it leads to slightly ($\lesssim 1\%$) underestimations comparing with FLAMINGO at $k\gtrsim 0.3\hmpc$ and $z=0.0$ for cosmology with $\sum m_{\nu}=0.24\,\mathrm{eV}$.}.
For convenient usage, we also provide the emulator of this transformation in our released package.
The emulation accuracy can reach 0.05\% for all scales and redshift ranges by emulating the HMCODE-2020 results under 513 training cosmologies.

\begin{figure}[H]
    \centering
    \includegraphics[width=0.4\textwidth]{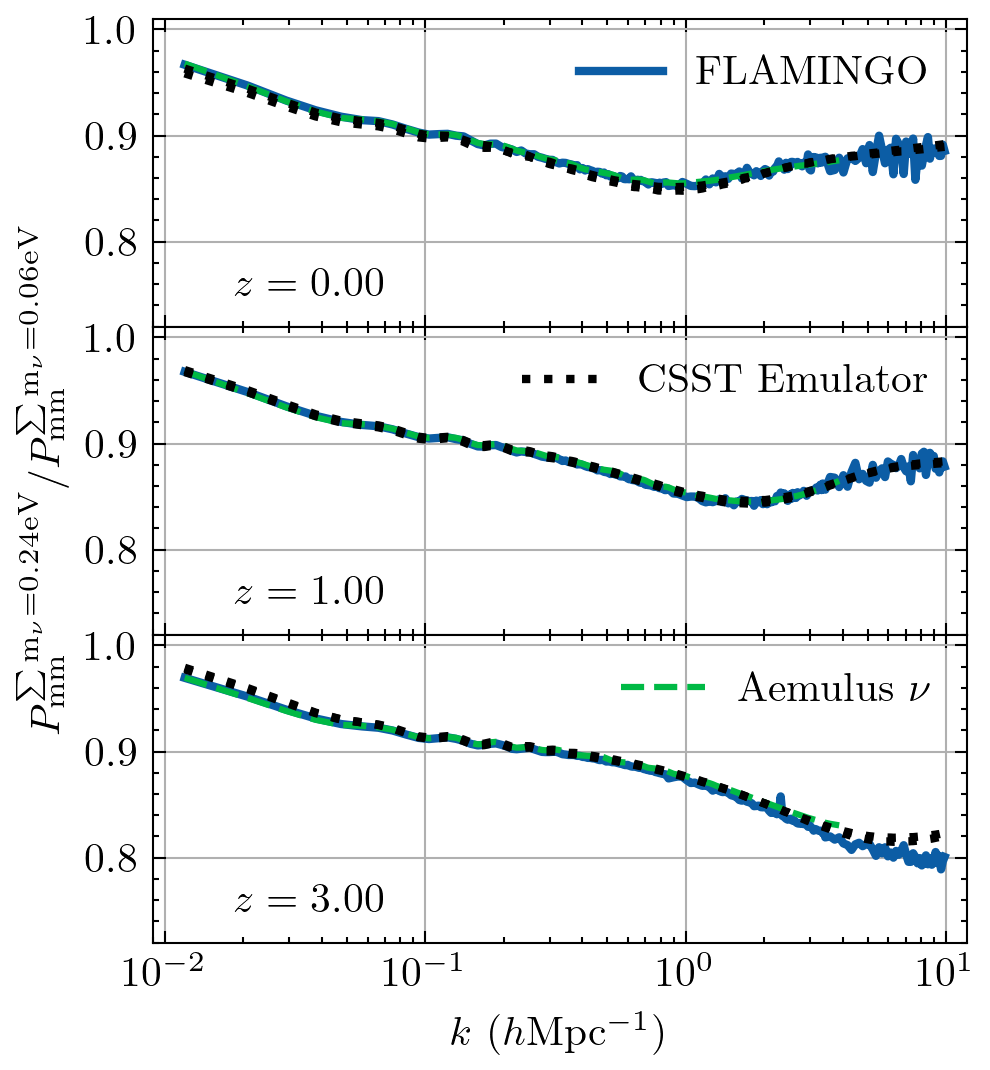}
    \caption{The ratio of the matter power spectrum from cosmology with $\sum m_{\nu}=0.24~\mathrm{eV}$ and the one with neutrino mass of $0.06~\mathrm{eV}$.
    The results of \textsc{FLAMINGO}, \texttt{CSST Emulator}, and \texttt{Aemulus-$\nu$} are shown in solid, dotted, and dashed lines, respectively.
    Different redshifts are illustrated in different panels.
    }
    \label{fig:emu_flamingo}
\end{figure}

The neutrino suppression measured from \textsc{FLAMINGO} simulations is illustrated by the blue solid lines in Fig.~\ref{fig:emu_flamingo}.
The spoon-like feature is derived from the ratio of the power spectra of the PlanckNu0p24Fix\_DMO and Planck\_DMO simulations.
The green dashed and black dotted lines represent the results from \texttt{Aemulus-$\nu$} and \texttt{CSST Emulator}, respectively.
A great agreement is observed for both emulators and simulations, which indicates our treatment of neutrinos is sufficiently accurate even when compared to the particle-based methods.
The slight offset at large scales is caused by the absence of the last step in the Newtonian motion gauge method.

\subsection{Comparison with Various Simulations}

The errors introduced by the emulation process have been validated to be comparable with errors arising from simulation configurations in Fig.~\ref{fig:pkemu-different-methods}.
It is also important to compare our prediction with data simulated by different codes in order to assess potential systematic uncertainties.
Several investigations have been conducted on cosmological simulation code comparisons (e.g., \cite{2014MNRAS.440..249S,2016JCAP...04..047S,2019MNRAS.485.3370G,2021MNRAS.506.2871S}).
However, it should be noted that there are no definitive correct results for the matter power spectrum in the extremely nonlinear regions.
This cross-validation serves primarily as a consistency check instead of quantifying the accuracy.

\begin{figure}[H]
    \centering
    \includegraphics[width=0.4\textwidth]{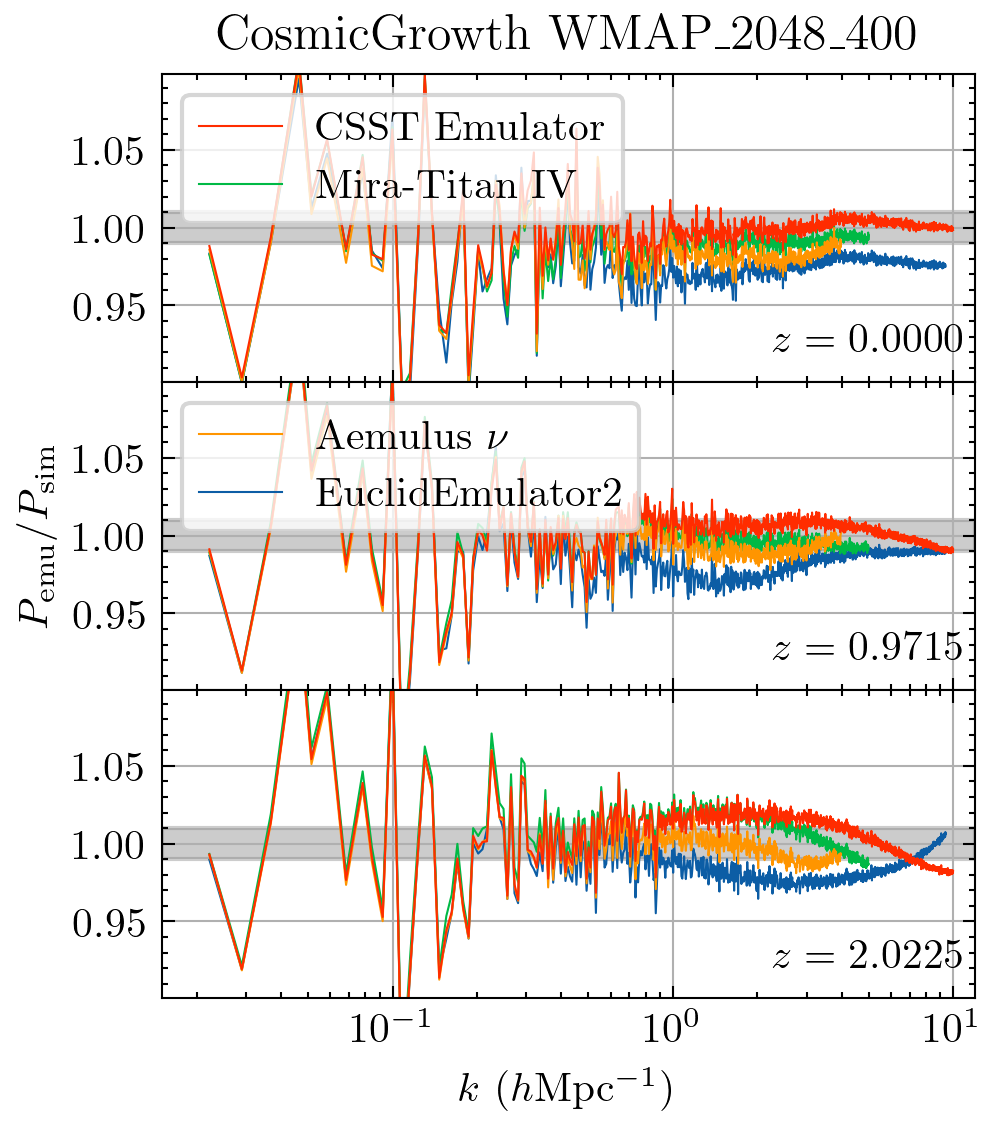}
    \caption{
    Comparison of matter power spectra of \textsc{CosmicGrowth} simulation with four emulator predictions. The results of $z=0.0$ to $z=2.0225$ are illustrated in top to bottom panels. 
    }
    \label{fig:emu_8511}
\end{figure}

Firstly, we compare the power spectrum predicted by the emulator with the WMAP\_2048\_400 result, one of the high-resolution simulations in \textsc{CosmicGrowth}, as shown in Fig.~\ref{fig:emu_8511}.
This simulation evolves $2048^3$ particles in a $400^3 \mpcht$ cubic box with a corresponding particle mass of $5.54 \times 10^8\,h^{-1}\Msun$, using a massively parallel $\mathrm{P^3M}$ $N$-body code.
The cosmology is listed in Table~1 in \cite{2019SCPMA..6219511J}.
Three other emulators, \texttt{Mira-Titan IV}, \texttt{Aemulus-$\nu$}, and \texttt{EuclidEmulator2}, are also illustrated by solid lines with different colors.
The maximum wavenumber $k$ varies as the claimed minimum scale that each emulator provides.
The grey shaded region shows a 1\% difference.
We observe excellent agreement at $z=0.0$ and $z=0.9715$ for both \texttt{CSST Emulator} and  \texttt{Mira-Titan IV}.
At the highest redshift, the discrepancy is slightly larger for these two emulators at $k\sim 2\,\hmpc$.
The \texttt{Aemulus-$\nu$} slightly underestimates the power for $k\gtrsim 1\,\hmpc$, especially at $z=0.0 $.
This is probably related to the difference in the specification of training simulations, e.g., the initial condition and mass resolution.
For the \texttt{EuclidEmulator2}, however, there is a systematic $2\sim 3\%$ underestimation when comparing its prediction with the WMAP\_2048\_400 simulation across all redshifts.
Both the amplitude and trend with redshift are similar to the resolution-induced power suppression, illustrated in Figure~4 of \texttt{EuclidEmulator2} paper \cite{2021MNRAS.505.2840E}. 

\begin{figure}[H]
    \centering
    \includegraphics[width=0.4\textwidth]{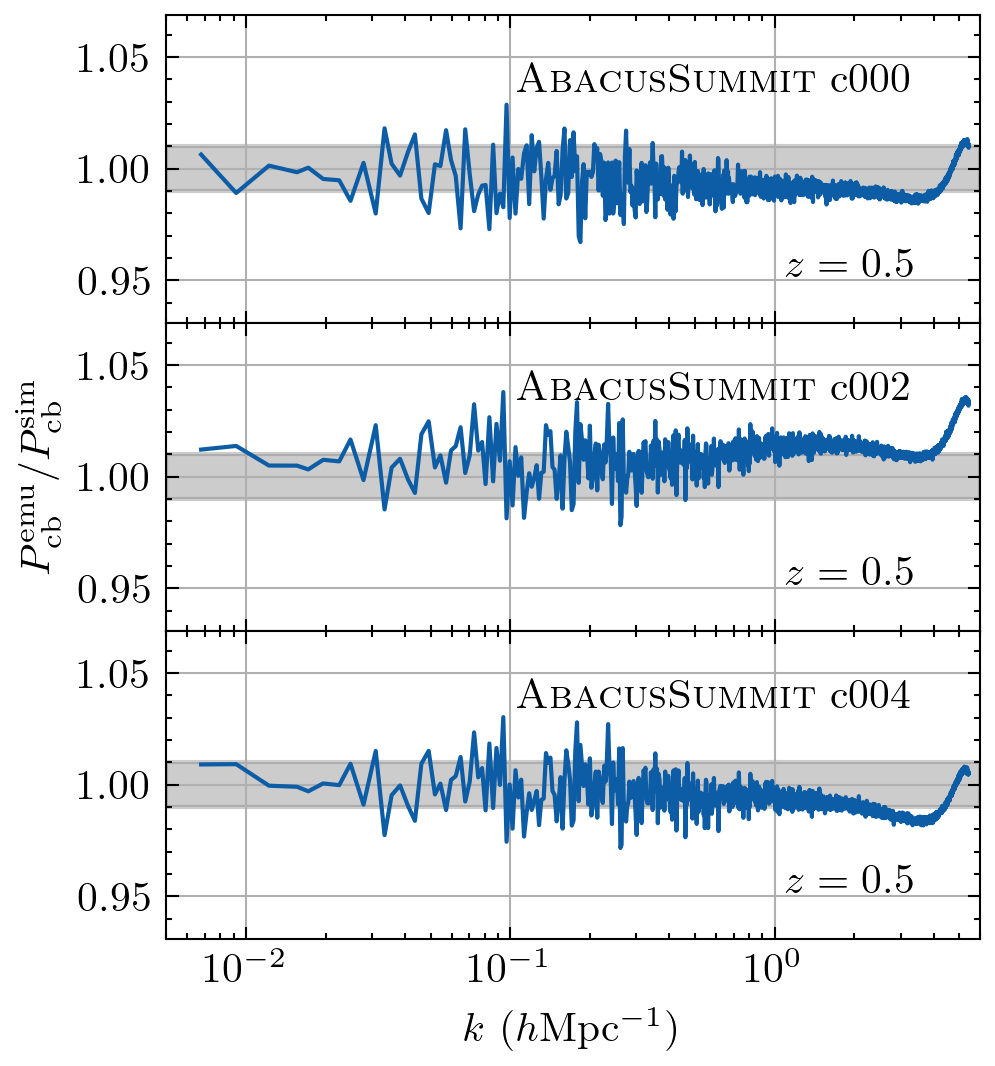}
    \caption{Comparison of cold-component power spectra of \textsc{AbacusSummit} simulations with \texttt{CSST Emulator} for three different cosmologies. We only show the ratios at $z=0.5$ for better visualization. }
    \label{fig:emu_AbacusSummit}
\end{figure}

The above comparison is validated only on the WMAP cosmology.
Next, we employ three `Fixedbase' simulations of the recent \textsc{AbacusSummit} suite generated by the Abacus $N$-body code \cite{2021MNRAS.508..575G}.
The `Fixedbase' series simulates $4096^3$ particles in a volume of $1.18^3\,\gpcht$ with the fixed-amplitude initial condition.
We choose three different cosmologies: the fiducial Planck 2018 cosmology including one single $0.06~\mathrm{eV}$ massive neutrino (\textsc{AbacusSummit} c000), the $w_0 w_a \mathrm{CDM}$ model with $w_{0} = -0.7$, $w_{a} = -0.5$ (\textsc{AbacusSummit} c002), and the baseline cosmology with lower $\sigma_8=0.75$ (\textsc{AbacusSummit} c004).
The ratios of the \texttt{CSST Emulator} predictions to simulation results for the cb-component power spectra are demonstrated in Fig.~\ref{fig:emu_AbacusSummit}.
Here, we do not compare with the \texttt{EuclidEmulator2} results because it only predicts the nonlinear power of total matter.
The \texttt{Aemulus-$\nu$} is not illustrated due to the absence of $w_{a}$ in its cosmological parameter space.
Agreements within 1\% are observed for all cosmologies at $z=0.5$ for $k\lesssim 3\,\hmpc$.
The upwarp towards the Nyquist frequency is likely due to the incomplete correction for the aliasing effect in the power spectrum provided by \textsc{AbacusSummit} website\footnote{\url{https://lgarrison.github.io/AbacusCosmos/data_specifications}} (see the right panel in Fig.~2 of \cite{2024JCAP...09..044W}).
Additionally, residual cosmic variance for the fixed-amplitude-only method leads to slightly larger fluctuations at the BAO scales.

In conclusion, our \texttt{CSST Emulator} can predict the matter power spectrum from higher resolution simulations obtained from other $N$-body codes, with $\sim 1\%$ accuracy at $z\leq 2.0$ and $k \lesssim 10\,\hmpc$.

\subsection{Comparison with Other Emulators}

\begin{figure*}
    \centering
    \includegraphics[width=0.9\textwidth]{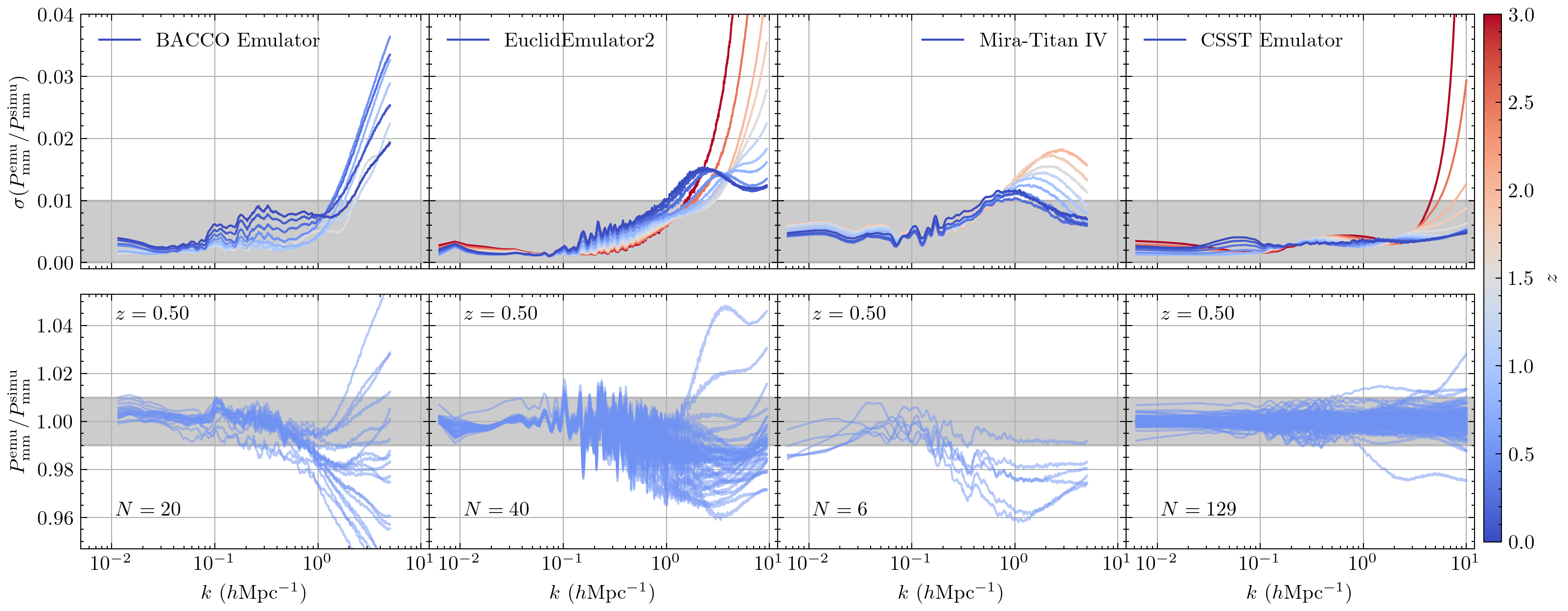}
    \caption{The top panels show the 68th percentile errors for total matter power spectrum $P_\mathrm{mm}(k,z)$ of \texttt{BACCO}, \texttt{EuclidEmulator2},  \texttt{Mira-Titan IV} and \texttt{CSST Emulator} (from left to right column).
    The redshifts from $z=3.0$ to $z=0.0$ are represented by color from red to blue.
    The ratio of the emulated power spectrum to the \textsc{Kun} simulations at $z=0.50$ is exhibited in the lower panel.
    The grey region shows the 1\% difference.
    The number of simulations used to validate the errors, i.e., the number of lines in the plot, is shown on the lower left of each subplot.
    }
    \label{fig:emus-compare}
\end{figure*}

The validity of the power spectrum emulation has been proved in previous sections.
In this part, we utilize the spectra measured from our training simulations to evaluate three other emulators and compare their performance with the LOO errors of the \texttt{CSST Emulator}.
The top panel in Fig.~\ref{fig:emus-compare} summarizes the performance of \texttt{BACCO}, \texttt{EuclidEmulator2},  \texttt{Mira-Titan IV} and \texttt{CSST Emulator} for $P_\mathrm{mm}(k,z)$.
The corresponding ratios between emulations and individual simulation results at $z=0.50$ are illustrated in the lower panel.
The chosen redshift is close to the effective redshift of the photometric catalog for the upcoming CSST survey \cite{2024MNRAS.527.5206Y}.
Note that the number of simulations to validate the 68th percentile errors varies for each emulator due to the different parameter space coverage.
This information is texted at the lower left in the bottom panel.
We caution that the comparison between different emulators is not entirely fair as \texttt{CSST Emulator} is trained on the test simulation suite itself.
Therefore, the superior convergence of the \texttt{CSST Emulator} is expected, as it is free from the effects of simulation setups.

The \texttt{BACCO} and \texttt{Mira-Titan IV} emulators predict the power spectrum only up to $k\leq 5\,\hmpc$ for $z\leq 1.5$ and $z\leq 2.02$, respectively.
The relative errors of the \texttt{BACCO} emulator are $2\sim4\%$ for $k\geq 1\,\hmpc$, which is consistent with Figure~16 in \cite{2021MNRAS.505.2840E}, in which the power spectra of \texttt{EuclidEmulator2} were corrected by the resolution correction factor (RCF).
However, the output of the released \texttt{EuclidEmulator2} omits this RCF.
The 68th error of the  \texttt{Mira-Titan IV} emulator is relatively small ($\leq 2\%$) compared to those of \texttt{BACCO} and \texttt{EuclidEmulator2}.
While it should be noted that this result is based on only six test simulations from the \textsc{Kun} suite, whose cosmologies are closer to the center of the full training parameter space.
The bias at small scales is potentially explained by the different mass resolutions of the training data.
In the middle left panel, the \texttt{EuclidEmulator2} exhibits a $\sim 2\%$ discrepancy at low redshifts compared to the simulation results, while becoming worse with $z$ and $k$ increased.
The significant bias at the highest redshift and nonlinear scales is caused by the dominant shot noise.
The suppression of moderate redshift observed in the ratio of emulator to simulation is similar to the discovery in Fig.~\ref{fig:emu_8511}.
We expect that this inaccuracy may be omitted if the resolution-induced correction is included in the package of \texttt{EuclidEmulator2}.

For the \texttt{CSST Emulator} outlined in this work, the 68th percentile error at $z\leq2.0$ has met the demanded accuracy for the Stage-\Rmnum{4} surveys.
The slightly worse performance at higher redshifts originates from the noise in the simulation itself.
We discuss this issue in detail in \ref{sec:shotnoise}.
We observe excellent agreement ($\leq 1\%$) at $z=0.5$ for most cosmologies.
The worst prediction ($P_\mathrm{emu}/P_\mathrm{sim} < 0.98$) occurs for only one cosmology, which lies at the left corner of the whole 8D cosmological parameter space.
Therefore, we conclude that our \texttt{CSST Emulator} can predict the power spectra at $k\leq 10\,\hmpc$ for $z\leq 2.0$ with approximately $1\%$ accuracy, except for regions near the boundaries of the parameter space.
For the $2.0< z\leq3.0$, the accuracy remains adequate at $k\leq5\,\hmpc$.
However, shot noise becomes dominant for some cosmologies with lower $\sigma_8(z=3.0)$ if we go to smaller scales.

\label{sec:comp_other_emulators}

\section{Conclusion and Discussions}
\label{sec:conc}

With the development of modern cosmology, Stage-\Rmnum{4} surveys have unprecedented statistical power to constrain cosmological parameters, including the nature of dark energy, the mass sum of neutrinos, among others.
It is essential to develop accurate nonlinear prediction tools to control the systematics from the theoretical side to the same percentage level.
This work, as the first one of a series, aims to construct a cosmological emulator to predict the fundamental statistic, the matter power spectrum, for the planed CSST survey.

The emulator is named \texttt{CSST Emulator}, and is constructed on the new \textsc{Kun} simulation suite.
It contains 129 high-resolution simulations with the cosmological model sampled by the Sobol sequence in an 8D parameter space. 
We modify the Gadget-4 to implement the Newtonian motion gauge method to simulate the neutrino effect.

The emulation is based on the ratio of the nonlinear power spectrum to the HMCODE-2020 prediction.
For each cosmology, two extra FastPM runs and a smoothing procedure are adopted to overcome the residual cosmic variance.
The performance of the emulator is validated through leave-one-out error analysis, as well as by comparison with external simulations and emulators.
We conclude that \texttt{CSST Emulator} can achieve $1\%$ accuracy $z \leq2.0$ at $k\leq 10\,\hmpc$, and $2\%$ for $2.0\leq z \leq3.0$ and $k\leq6\,\hmpc$.
The $\sim 1\%$ accuracy over the wide parameter space has met the requirement for the CSST imaging and spectroscopic observations.

The \texttt{CSST Emulator} is publicly available at \url{https://github.com/czymh/csstemu}.
This tool is a user-friendly Python code and can output the nonlinear power spectrum for $k\leq 10\,\hmpc$ and $0.0\leq z \leq3.0$, requiring only \textsc{numpy} \cite{2020Natur.585..357H} and \textsc{scipy} \cite{2020SciPy-NMeth}.

Compared with previous studies, we conclude that the excellent performance of \texttt{CSST Emulator} is primarily attributed to the following improvements:
\begin{itemize}
    \item We compare the influence of different high-dimensional sampling techniques for the first time.
    The well-performed Sobol sequence sampling is used to generate the training set with potential extensions on both the number and the dimension.
    \item The cosmic variance is significantly mitigated by the fixed \& paired technique, and the computational cost is well controlled by using the FastPM runs as the matched and paired simulations.
    By further applying a Savitsky-Golay filter, the fluctuations at BAO scales are effectively suppressed to well below $1\%$.
    \item The improvement by incorporating the well-established theory into the emulation process is tested by comparing three different strategies.
    The successful HMCODE-2020 is employed to provide prior information in the emulator for the first time, ensuring that the errors introduced during the emulation are less than $1\%$.
    \item The extra emulation on the HaloFit and HMCODE-2020 predictions enables the \texttt{CSST Emulator} predict spectra per cosmology within $\lesssim 15\,\mathrm{ms}$, and independent of external packages.
\end{itemize}

The main usage of the emulator is to fully capture the cosmological information of galaxy surveys at small scales, enabling tighter constraints.
There are typically three roadways to provide accurate model predictions of galaxy clustering or weak lensing observations.
Firstly, we can directly construct the emulator of galaxy two-point statistics through 
the galaxy catalog generated by employing the HOD or SHAM to the halo catalog (e.g., \texttt{CosmicEmu} \cite{2015ApJ...810...35K}, \texttt{Aemulus} \cite{2019ApJ...874...95Z,2023ApJ...948...99Z,2024ApJ...961..208S}, \texttt{Abacus} \cite{2020MNRAS.492.2872W}, \texttt{AbacusSummit} \cite{2022MNRAS.515..871Y}).
The major limitation is that the halo-galaxy connection parameterization must be determined, indicating that the emulator is constructed for the specific galaxy samples.
However, this approach facilitates the extensions for other high-order statistics.
The second approach is to emulate the halo mass function and halo clustering for different mass thresholds first, and then to predict the observational probes by incorporating theoretical halo-galaxy connections (e.g., \texttt{Dark Quest} \cite{2019ApJ...884...29N,2020PhRvD.102f3504K}).
The implementation is flexible because it focuses solely on the emulation of the halo statistics.
The clustering of any galaxy sample can be predicted by reasonably modifying the halo-galaxy connection formalism.
The third solution is to emulate the basis functions of the bias expansion model (e.g.,
\texttt{AbacusSummit} \cite{2021JCAP...09..020H},
\texttt{BACCO} \cite{2023MNRAS.524.2407Z,2023MNRAS.520.3725P},
\texttt{Aemulus-$\nu$} \cite{2023JCAP...07..054D}).
This implementation does not even rely on halo-finding algorithms.
We plan to provide multiple emulators with different schemes to take full advantage of the 3.1 PB data in \textsc{Kun} simulation suite.

Another natural extension of this work is to construct a series of emulators for various useful statistics, particularly high-order statistics.
Due to the non-Gaussianity of the late-time Universe, it is essential to combine the high-order statistics, e.g., peak, void, Minkowski functionals, and scattering transform, to break the degeneracy between cosmological parameter constraints.
However, modeling these nonlinear quantities is challenging from first principles.
Our simulations can directly emulate them to exploit additional cosmological information in the Bayesian analysis.
Besides, it is possible to construct field-level emulators to extract the maximum amount of constraint power (e.g., \cite{2024MNRAS.529...89P,2024MNRAS.535.1258D}).
Recent weak lensing surveys have demonstrated the significant potential of this approach (e.g., KiDS-1000 \cite{2022PhRvD.105h3518F}, DES-Y3 \cite{2022MNRAS.511.2075Z,2023MNRAS.525..761Z,2025MNRAS.536.1303J,2024arXiv240510881G}).

Furthermore, universal laws play a crucial role in cosmology. 
They not only deepen our understanding of the underlying principles governing the universe but also simplify the analysis of cosmological data. 
For example, several studies have explored the halo mass function and developed accurate mathematical expressions to describe it (e.g., \cite{2008ApJ...688..709T,2014JCAP...02..049C,2022MNRAS.509.6077O}). 
It would be interesting to utilize the \textsc{Kun} simulation suite to search for and validate potential universality in the halo mass function or other functions.
Some recent studies have also explored alternative definitions (e.g., splashback radius \cite{2016MNRAS.459.3711S,2020ApJ...903...87D}, depletion radius \cite{2021MNRAS.503.4250F,2023ApJ...953...37G}) of the halo boundary, which could help provide more physical and complete descriptions of structure formation.
In particular, Zhou \& Han \cite{2023MNRAS.525.2489Z,2025ApJ...979...55Z} have demonstrated that a new halo model based on the depletion radius is capable of accurately predicting multiple statistics of the halo and matter fields.
It would also be interesting to explore these new developments of the halo model using the \textsc{Kun} simulations and incorporate them into our emulator for more versatile and accurate predictions in future studies.

\Acknowledgements{
We thank Pengjie Zhang, Ji Yao, Ming Li, and Christian Fidler for the useful discussions.
This work was supported by the National Key R\&D Program of China (No. 2023YFA1607800, 2023YFA1607801, 2023YFA1607802), the National Science Foundation of China (Grant Nos. 12273020, 12133006), the China Manned Space Project with No. CMS-CSST-2021-A03, the ``111'' Project of the Ministry of Education under grant No. B20019,
and the sponsorship from Yangyang Development Fund.
The \textsc{Kun} simulation suite is run on 
Kunshan Computing Center.
The analysis is performed on the Gravity
Supercomputer at the Department of Astronomy, Shanghai Jiao Tong University, and the $\pi \,2.0$ cluster supported by the Center for High Performance Computing at Shanghai Jiao Tong University.
}

\InterestConflict{The authors declare that they have no conflict of interest.}


\bibliographystyle{scpma}
\bibliography{biblio.bib}

\begin{appendix}
\section{ Sobol Sequence }
\label{sec:sobol}

Sobol sequences are deterministic, quasi-random low-discrepancy sequences designed for efficient high-dimensional numerical integration and Monte Carlo simulations.
The sampling points are generated through binary bitwise operations and the $2^{N}$ coordinates in each dimension belong to $\{ \frac{k}{2^N},\, k\in[ 0,1,...,2^{N}-1  ] \}$.
While each dimension uses a unique generator matrix to reorder the coordinates.
A simple example for 8 points in 3D space is shown in Tab. \ref{tab:1}.
When increasing the sample size, points sequentially fill space using base-2 stratification, maintaining uniform density in the whole space ($2^{N}$ points achieve perfect $2^{N}$-box stratification).
This process is independent of the chosen number and the samples' dimension, indicating the extendibility in the experimental design.
Lower-dimensional views (e.g., 2D planes) may exhibit grid-like structures due to deterministic bit alignment, though global uniformity persists in high dimensions.
Compared to pseudo-random sequences, Sobol points better avoid large gaps/clusters by simultaneously enforcing exponentially spaced coordinate permutations in all dimensions.
Unlike Latin hypercube sampling, which struggles with high-dimensional stratification, the sequence maintains its uniformity properties as the dimension increases.

\begin{table}[H]
\centering
\caption{An example of Sobol sequence sampling. }
\begin{tabular}{|c|c|c|}
\hline
Dimension 1 & Dimension 2 & Dimension 3 \\ 
\hline
0 & 0 & 0 \\
\hline
0.5 & 0.5 & 0.5 \\
\hline
0.75 & 0.25 & 0.25 \\
\hline
0.25 & 0.75 & 0.75 \\
\hline
0.375 & 0.375 & 0.625 \\
\hline
0.875 & 0.875 & 0.125 \\
\hline
0.625 & 0.125 & 0.875 \\
\hline
0.125 & 0.625 & 0.375 \\
\hline
\end{tabular}
\label{tab:1}
\end{table}

\section{ Shot Noise Contamination }
\label{sec:shotnoise}

\begin{figure}[H]
    \centering
    \includegraphics[width=0.4\textwidth]{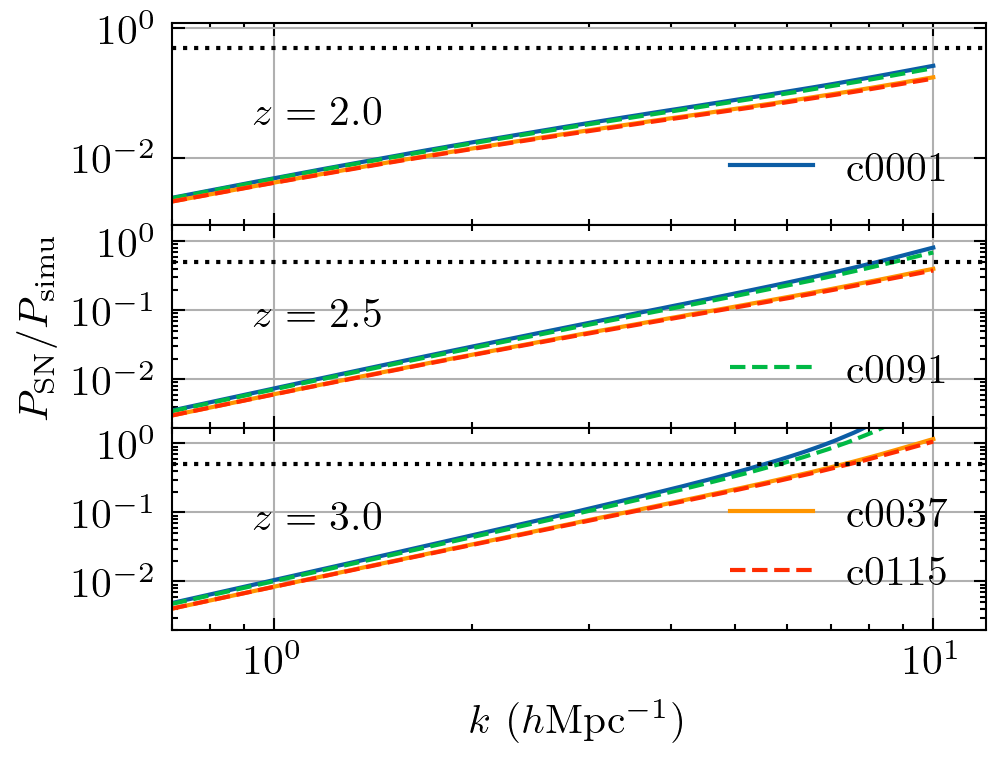}
    \caption{Ratios between shot noise and power spectra for four least clustering cosmologies at $2.0 \leq z \leq 3.0$.
    }
    \label{fig:shotnoise}
\end{figure}

\begin{figure}[H]
    \centering
    \includegraphics[width=0.4\textwidth]{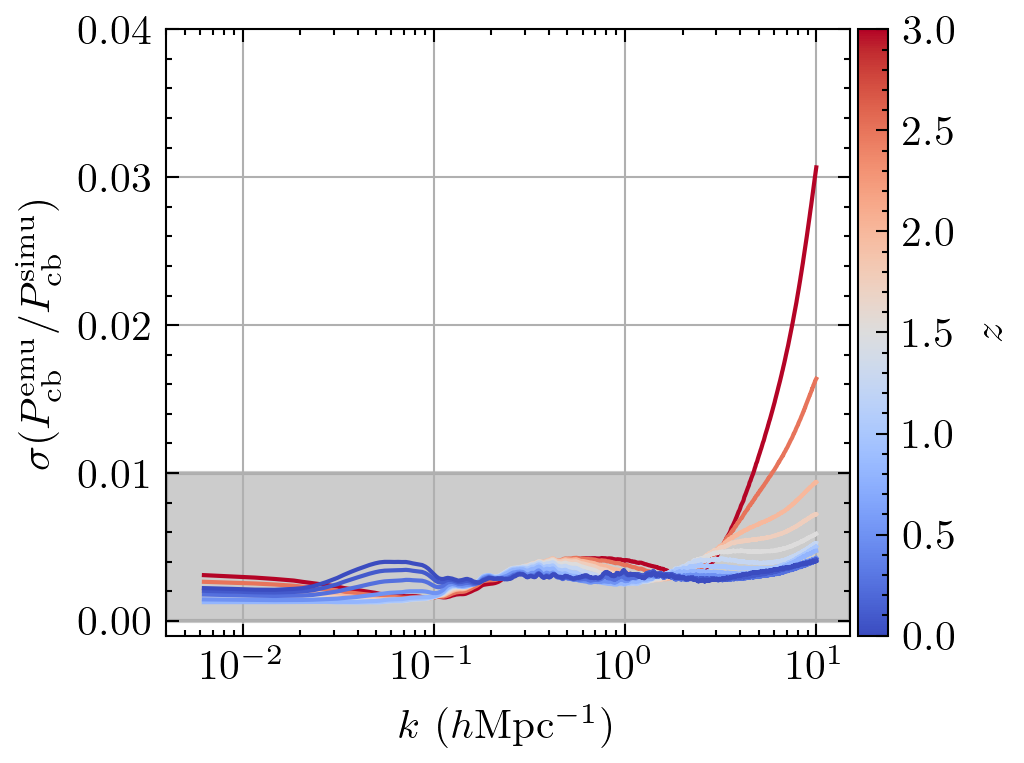}
    \caption{Same as the right panel in Fig.~\ref{fig:pkemu-different-methods}, but  four cosmologies (c0001, c0037, c0091 and c0115) are removed in evaluating the 68th percentile errors.}
    \label{fig:pk-emu-remove4cosmos}
\end{figure}

In this appendix, we investigate the influence of the shot noise on the power spectrum at high redshifts.
The ratios of shot noise to the power spectra for the three highest redshifts are illustrated in Fig.~\ref{fig:shotnoise}.
The four cosmologies selected have the smallest $\sigma_8(z=3.0)$, and their parameters are detailed on \footnote{\url{https://csst-emulator.readthedocs.io/en/dev/cosmologies.html}}.
The black dotted lines represent $P_\mathrm{SN}/P_\mathrm{simu} = 0.5$, indicating that the shot noise is non-negligible.
We see significant shot noise at $z=2.5$ and $3.0$ for four cosmologies.
If we maintain the emulation process the same as the main text but exclude these samples when calculating the 68th percentile LOO errors, the result is shown in Fig.~\ref{fig:pk-emu-remove4cosmos}.
The large errors at higher redshifts are significantly suppressed.
This indicates the mass resolution of $2.87 \frac{\Omega_\mathrm{cb}}{0.3} \times 10^{9}\, h^{-1}\Msun$ is inadequate, leading to a slight degradation in the performance of our emulator for $z>2.0$.
Fortunately, these cosmologies lie in the outer regions of the entire 8D parameter space.
This means that the emulator precision in the inner region is sufficiently accurate.
We demonstrate the detailed error maps in \ref{sec:error_map}.

\section{ Emulator Performance Map }
\label{sec:error_map}

\begin{figure}[H]
    \centering
    \includegraphics[width=0.4\textwidth]{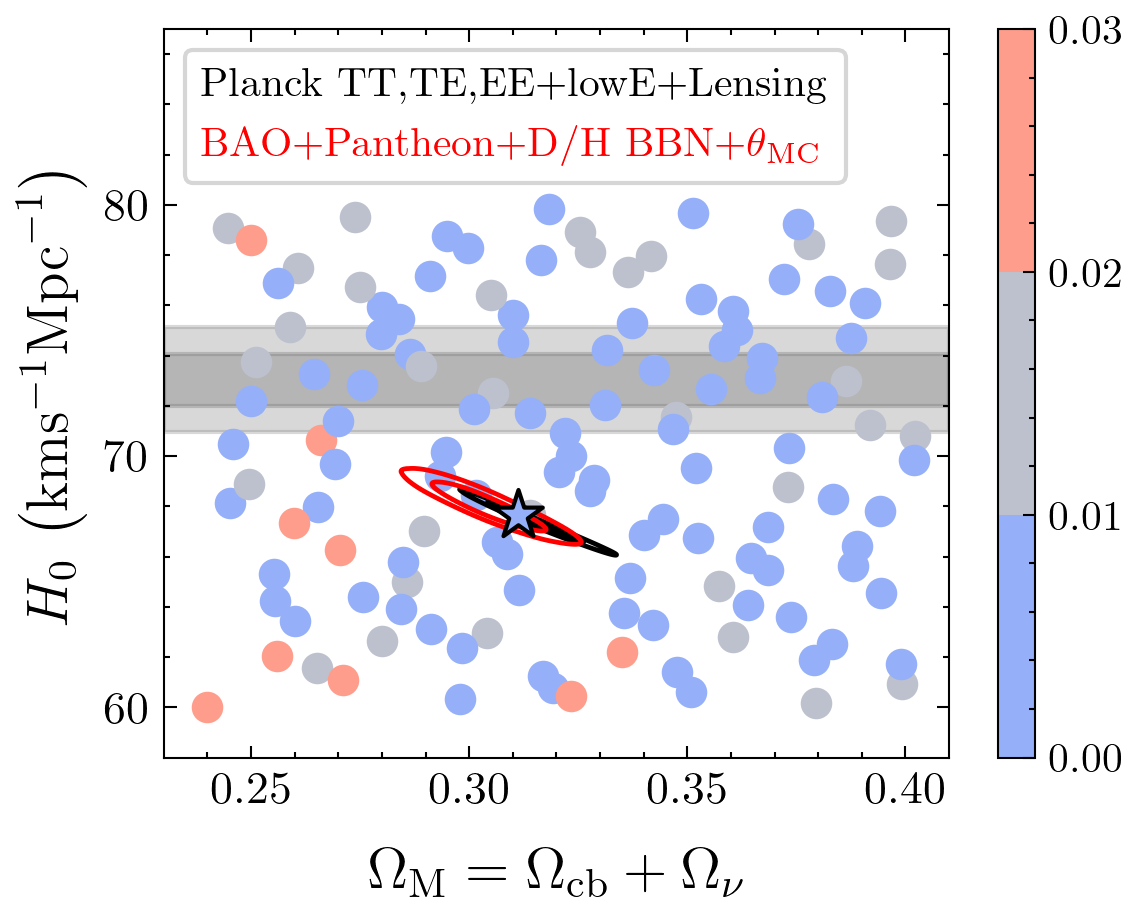}
    \caption{The Error map for all 129 cosmologies at the $\Omega_\mathrm{M} \mbox{-} H_\mathrm{0}$ plane. The points with large errors ($\epsilon > 2\%$) only appear in the lower left corner. The constraint from Planck CMB data \cite{2020A&A...641A...6P} is illustrated as the black solid contour.
    The red contour represents the constraint from the combination of BAO+Pantheon+D/H BBN+$\theta_\mathrm{MC}$ \cite{2018ApJ...859..101S,2018ApJ...855..102C}.
    The grey bands represent the local distance-ladder measurement of Riess et al.~2021 \cite{2022ApJ...934L...7R}.
    }
    \label{fig:H0-OmegaM}
\end{figure}

To address the distribution of errors in the whole parameter space, we define the maximum error for each cosmology as follows:
\begin{equation}
\label{eq:err_max}
\epsilon=\max\left(\left| \frac{P^{\mathrm{emu}}_\mathrm{cb}(k, z)-P^{\mathrm{simu}}_\mathrm{cb}(k, z)}{P^{\mathrm{simu}}_\mathrm{cb}(k, z)}\right|\right),
\end{equation}
where $0.00628\,\hmpc \leq k \leq 10.0\,\hmpc$ and $z\in [0.0, 2.0]$.
The maximum errors of all training simulations on the $\Omega_\mathrm{M} \mbox{-} H_\mathrm{0}$ plane are shown in Fig.~\ref{fig:H0-OmegaM}, where the color scale indicates the magnitude of the error.
We also display the contour of parameter constraints from Planck CMB and BAO data from the Planck 2018 paper \cite{2020A&A...641A...6P} for comparison\footnote{Chains are downloaded from \url{https://pla.esac.esa.int}.}.
The measurements from the SNIa and Cepheid observations are shown by the grey bands \cite{2022ApJ...934L...7R}.
The different values of $H_\mathrm{0}$ determinations from both late-time distance ladder and CMB measurements are included in our parameter space.
We observe that the points with large errors are located far from the fiducial Planck cosmology, indicating that the accuracy in the vicinity of the fiducial cosmology is better than $1\%$.

\begin{figure}[H]
    \centering
    \includegraphics[width=0.4\textwidth]{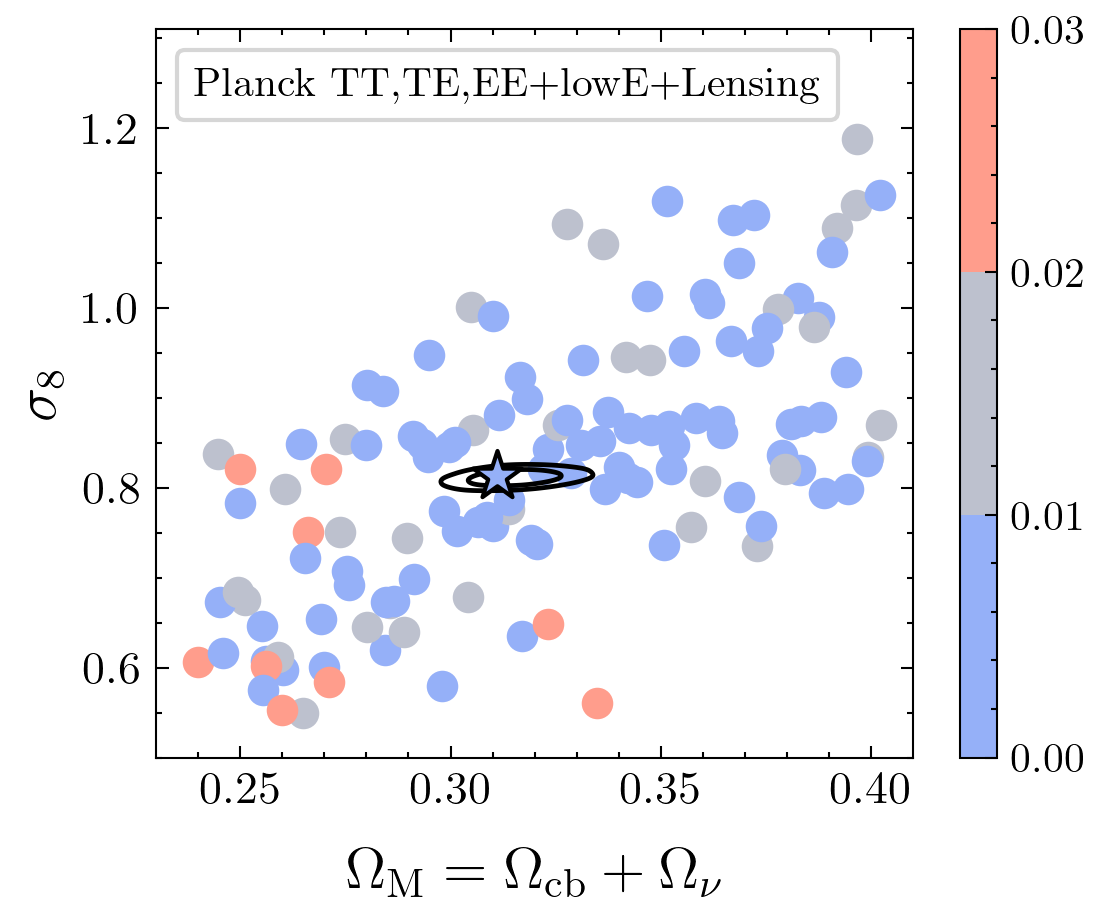}
    \caption{Similar with Fig.~\ref{fig:H0-OmegaM}, but on the $\Omega_\mathrm{M} \mbox{-} \sigma_8$ coordinate.}
    \label{fig:sigma8-OmegaM}
\end{figure}

For the weak gravitational lensing survey, the correlation signal is particularly sensitive to the amplitude of clustering and the total matter density at present.
Thus, we also show the error maps on the $\Omega_\mathrm{M} \mbox{-} \sigma_8$ plane in Fig.~\ref{fig:sigma8-OmegaM}, along with the Planck 2018 result for comparison.
Similar to the previous finding, our emulator achieves higher accuracy for cosmologies that are close to the current cosmological constraints.

\end{appendix}

\end{multicols}
\end{document}